\PassOptionsToPackage{table,dvipsnames}{xcolor}
\documentclass[sigconf]{acmart}

\usepackage{amsmath,amssymb,amsfonts}
\usepackage{graphicx}
\usepackage{textcomp}

\usepackage{graphicx} % Required for inserting images

\usepackage{algorithm}
\usepackage{algorithmic}
\usepackage{multirow}
\usepackage{enumitem}
\usepackage{listings}
\usepackage{xspace}

\usepackage{booktabs}
\usepackage{subfig}
\usepackage{mdframed}

\usepackage{tikz}

\usepackage{mdframed}
\newcommand{\finding}[1]{
    \begin{mdframed}[linecolor=gray,roundcorner=12pt,backgroundcolor=gray!15,linewidth=3pt,innerleftmargin=2pt, leftmargin=0cm,rightmargin=0cm,topline=false,bottomline=false,rightline = false]
        #1
    \end{mdframed}
}
\usepackage{makecell}
\usepackage{hyperref}

\AtBeginDocument{%
  }

\setcopyright{acmlicensed}
\copyrightyear{2018}
\acmYear{2018}
\acmDOI{XXXXXXX.XXXXXXX}
\acmConference[Conference acronym 'XX]{Make sure to enter the correct
  conference title from your rights confirmation email}{June 03--05,
  2018}{Woodstock, NY}
\acmISBN{978-1-4503-XXXX-X/2018/06}

\begin{document}

\title{LLM as an Execution Estimator: Recovering Missing Dependency for Practical Time-travelling Debugging}

\author{Yunrui Pei}
\affiliation{%
  \institution{Shanghai Jiao Tong University}
  \city{Shanghai}
  \country{China}
}
\email{yunruipei@sjtu.edu.cn}
\authornotemark[1]
% \authornote{the authors equally contribute to the paper}

\author{Hongshu Wang}
\affiliation{%
  \institution{National University of Singapore}
  \city{Singapore}
  \country{Singapore}
}
\email{hongshu.wang@u.nus.edu}
\authornotemark[1]

\author{Wenjie Zhang}
\affiliation{%
 \institution{National University of Singapore}
  \city{Singapore}
  \country{Singapore}
}
\email{wjzhang@nus.edu.sg}
\authornote{The authors equally contribute to the paper}

\author{Weiyu Kong}
\affiliation{%
  \institution{Shanghai Jiao Tong University}
  \city{Shanghai}
  \country{China}
}
\email{kwy160034@sjtu.edu.cn}

\author{Yun Lin}
\affiliation{%
  \institution{Shanghai Jiao Tong University}
  \city{Shanghai}
  \country{China}
}
\email{lin_yun@sjtu.edu.cn}
\authornote{Corresponding author}

\author{Jin Song Dong}
\affiliation{%
 \institution{National University of Singapore}
  \city{Singapore}
  \country{Singapore}
}
\email{dcsdjs@nus.edu.sg}

\renewcommand{\shortauthors}{Trovato et al.}


\newcommand{\linyun}[1]{{\textcolor{red}{LY: #1}}}
\newcommand{\wenjie}[1]{{\textcolor{ForestGreen}{wenjie: #1}}}
\newcommand{\hongshu}[1]{{\textcolor{blue}{hongshu: #1}}}
\newcommand{\yr}[1]{{\textcolor{orange}{yr: #1}}}

\lstset{
    language=Java,                % Set the language to Java
    basicstyle=\ttfamily\small,   % Use a monospaced font, small size
    keywordstyle=\color{blue}\bfseries, % Keywords in blue and bold
    stringstyle=\color{red},      % Strings in red
    commentstyle=\color{green!50!black}\itshape, % Comments in green and italic
    numbers=left,                 % Line numbers on the left
    numberstyle=\tiny\color{gray},% Line numbers are tiny and gray
    stepnumber=1,                 % Number every line
    numbersep=5pt,                % Distance between line numbers and code
    showspaces=false,             % Don’t show spaces as special characters
    showstringspaces=false,       % Don’t show spaces in strings
    showtabs=false,               % Don’t show tabs as special characters
    frame=single,                 % Add a frame around the code
    tabsize=4,                    % Set tab size to 4 spaces
    breaklines=true,              % Break long lines
    breakatwhitespace=true,       % Break only at whitespace
    captionpos=b,                 % Caption below the listing
    escapeinside={(*@}{@*)},      % Escape to LaTeX inside (*@ and @*)
    xleftmargin=10pt,  % Left margin
    xrightmargin=10pt, % Right margin
}

\newcommand{\tool}{\textsf{RecovSlicing}\xspace}
\newcommand{\debuggingtool}{\textsf{Tregression}\xspace}

% Overall benchmark data
\newcommand{\datasetCount}{ XX }
\newcommand{\totalSliceCount}{ 5986 }
\newcommand{\overallBenchmarkSize}{ xxx }

% RQ1 & RQ3 benchmark data
\newcommand{\NDSlicerBenchmarkSize}{905} % ND-Slicer
\newcommand{\LLMSlicerBenchmarkSize}{100} % LLM-Slicer
\newcommand{\DefectsBenchmarkSize}{ xxx } % Defects4J
\newcommand{\SlicerJBenchmarkSize}{ xxx } % Slicer4J (not used)
\newcommand{\GeneratedBenchmarkSize}{ 1170 } % Generated benchmark

\newcommand{\llmGeneratedDataset}{Hierarchical Synthetic Dataset\xspace}
\newcommand{\llmGeneratedSimple}{\textit{Basic Operations\xspace}}
\newcommand{\llmGeneratedComplex}{\textit{Noisy Context\xspace}}
\newcommand{\llmGeneratedAlias}{\textit{Variable Aliasing\xspace}}
\newcommand{\llmGeneratedMultiMethods}{\textit{Interprocedural\xspace}}
\newcommand{\llmGeneratedMultiFiles}{\textit{Inter-file\xspace}}

% RQ2 benchmark data
\newcommand{\RQIIBenchmarkSize}{ xxx }

% RQ4 benchmark data
\newcommand{\RQIVBenchmarkSize}{ xxx }

% RQ1 results

% RQ2 results

% RQ3 results

% RQ4 results
\newcommand{\RQIVSuccessRate}{89\%}
\newcommand{\RQIVImprovementPercentage}{16\%}

\begin{abstract}
  Dynamic data dependency, answering the question of ``\textit{why the variable has this value?}'',
  is important in program debugging.
  Given a program step $s$ reading a variable $v$,
  computing its dynamic data dependency (i.e., finding the step defining $v$) is non-trivial,
  which usually requires either
  %(1) alias inference for the variable definition,
  (1) exhaustive instrumentation for all the possible definitions of $v$ for one program run or
  (2) replicating the program run to re-investigate the reads and writes of $v$ for multiple times.
  In practice, the variable $v$ can be defined in a library call,
  which makes exhaustive instrumentation prohibitively expensive.
  Moreover, some programs can be non-deterministic,
  which makes replicating a program run infeasible.

  In this work, we propose \tool for computing dynamic data dependency 
  in a single run, with only partial instrumentation.
  %by our proposed \tool solution.
  %propose, \tool, an approach to interleaving LLM-based inference and program trace analysis for computing dynamic data dependency.
  We explore the intuition that
  LLM can potentially infer program dynamics based on a partially recorded trace and relevant code as its context.
  Given (1) a partially recorded trace of a program $P$ and
  (2) the slicing criteria consisting of a query step $s$ and a query variable $v$ read by $s$,
  \tool computes the runtime definition of $v$ on the trace by estimating the miss-recorded execution of $P$.
  In this work, we allow the user to specify implicit query variable,
  %such as the read variable inside a library call (e.g., 
  for example, the implicit library variable used in \texttt{list.get(i)}.
  %can either be explicit (e.g., the variable \texttt{a} in \texttt{Math.add(a, b)}) or implicit (e.g., the internal variable in \texttt{HashMap.get(key)}).
  %To this end, we infer miss-instrumented runtime variables and executions with an LLM-trace interleaving manner.
  Technically, built upon non-deterministic LLM, 
  we address the challenges of
  (1) \textit{precise} recovery of runtime variable value and structure from the recorded execution and
  (2) aligning the memory address of recovered variables and the recorded variables for definition analysis.
  %To this end, the inferred runtime behaviors include both runtime structure of variables and the program execution steps.
  %As for the miss-instrumented variables,
  %we adopt a novel adaptive in-context learning approach for \textit{precise} runtime structure and value of the query variable.
  %where the contexts fed to LLM are dynamically synthesized regarding the slicing criteria.
  %The recovered variables can also be used for further data dependency analysis.
  %As for the miss-instrumented executions,
  %we narrow down candidate steps, and
  %align the memory address of recovered variables with those recorded in the steps
  %for confirming the true variable-defining step.
  We extensively evaluate \tool against the state-of-the-art slicers such as Slicer4J, ND-Slicer, LLM Slicer, and re-execution Slicer on a total number of 8300 data-dependencies over 3 slicing benchmarks.
  The results show that \tool can significantly outperform the baselines. The accuracy and recall, achieving 80.3\%, 91.1\%, and 98.3\% on the three benchmarks, whereas the best baseline reaches 39.0\%, 82.0\%, and 59.9\% (accuracy), and 53.4\%, 79.1\%, and 87.1\% (recall), respectively.
  In addition, we integrate \tool in a dual-slicing based regression bug localizer,
  significantly improving its performance by locating 16\% more regressions.
\end{abstract}



% \keywords{program analysis, data dependency, LLM inference, in-context learning}

\received{20 February 2007}
\received[revised]{12 March 2009}
\received[accepted]{5 June 2009}

\maketitle

\section{Introduction}

Dynamic data dependency, answering why a variable $v$ is of value $val$,
is important for inferring the program causality, which can widely applied in program debugging
\cite{weiser1984program, korel1988dynamic, agrawal1990dynamic, venkatesh1991semantic, canfora1998conditioned, gupta1995hybrid}.
%\cite{weiser1984program}.
%Dynamic slicing techniques \cite{xx} are designed to identify all the program statements
%which is control or data depended by a given program statement during execution.
%When a program $P$ running upon an input $I$ to have a trace $\tau$
%where
%
%
%with a variable,
%we would like to identify the variable-defining step
%for a variable $v$ read by a step $s$,
%in its execution.
In practice, computing the dynamic data dependency of a runtime variable (i.e., when the variable is defined in the execution)
is non-trivial in the modern object-oriented programs.
Generally, we need to overcome the following challenges:
\begin{itemize}[leftmargin=*]
  \item {\textbf{Instrumentation Cost:}}
    Program instrumentation can be used to track the runtime definition of the variables.
    However, in modern program languages such as Java,
    the variable definition can happen in third-party or native library.
    Thus, instrumenting the definition can incur either large overhead
    (as the primitive library calls can happen everywhere in the subject program) or even not possible (as native library is non-instrumentable).
%  \item {\textbf{Alias Inference:}}
%    To selectively track the variable-defining program statements,
%    we usually need to infer variable alias,
%    which is particularly challenging for the modern object-oriented design
%    where objects can have complex reference relationships, polymorphism, and dynamic dispatch.
  \item {\textbf{Replication Feasibility/Cost:}}
    In addition, one can choose to re-execute the programs for multiple times to track the runtime definition of a variable,
    based on the assumption that rerunning a program is cheap and feasible.
    However, 
    a program could be non-deterministic (13\% of failed builds are due to flaky tests \cite{tahir2023test}).
    In addition, replicating an observed fault could be very expensive \cite{WANG2024107338,chaparro2019assessing,jin2012bugredux,rahman2022works,kang2024evaluating}. %\linyun{TODO, add more works on the cost of bug replication}.
    %Further, in some practical scenarios of computing data dependency,
    %we might not have the chance to re-instrument and re-execute the subject program,
    %which can happen for non-deterministic and concurrent programs.
\end{itemize}

Researchers have proposed various solutions to address the above challenges.
Slice4J \cite{ahmed2021slicer4j}, as a state-of-the-art tool integrating many traditional dynamic slicing techniques \cite{palepu2013improving, palepu2017dynamic},
applies full instrumentation on application code for computing data dependency.
While achieving leading performance in simple code,
its performance can be limited when parsing the data dependency cross external library calls.
For example, a library call \texttt{StringBuffer.\-app\-end()} indicates a variable definition and
a library call to \texttt{String\-Buffer.to\-String()} indicates a variable read,
which can form a data dependency.
However, instrumenting those JDK libraries can leads to unstable performance and unsafe results~\cite{huang2024revealing,lefort2021j}.
Despite Slice4J allows \textit{manual} library summary~\cite{palepu2013improving, palepu2017dynamic, ahmed2021slicer4j} to mitigate the issue,
the used libraries in practice can be extensive and new libraries are emerging,
it is less practical to manually summarize all the libraries.

With the emergence of LLM,
learning-based slicing techniques are proposed, such as ND-Slicer \cite{yadavally2024predictive} and LLM-Slicer \cite{shahandashti2024program},
for dynamic data dependency,
expecting LLM can parse code dynamics based on static code.
ND-Slicer is a predictive dependency analysis which trains an encoder-decoder transformer \cite{vaswani2017attention}
to predict the runtime dependencies from the input code tokens.
However, program dependency analysis is more of a logical deduction problem than of a statistical induction problem.
Training-based approaches could suffer from
(1) limited program knowledge such as unknown library design and
(2) distribution shift problem \cite{kuang2020stable, liang2023uncertainty} where the training dataset is not representative for all the practical scenarios.
In contrast, LLM-Slicer is a prompt-engineering based solution built upon the state-of-the-art LLMs such as GPT-4o \cite{islam2024gpt}, GPT-3.5 Turbo \cite{ye2023comprehensive}, and Llama-2 \cite{touvron2023llama}.
Nevertheless, LLMs can hallucinate its prediction especially when there is a big gap between the code and its unrolled execution.

In this work, we introduce \tool for precise dynamic data dependency by recovering the execution with LLM, which
only requires partial instrumentation of the subject program
%(2) mitigates alias-inference challenge, and
without re-execution and re-instrumentation.
Different from LLM-based solutions to infer dynamic dependency purely based on static code,
\tool infers the dependency by estimating the missing execution from the light-weighted execution.
%without suffering from the distribution shift.
%Our rationale lies in that LLM can be guided to \textit{execute} the static code upon its recorded context.
Given an input $I$ to the subject program $P$, 
we can have a trace $\tau$ based on the partially instrumented $P$.
Given a slicing criteria consisting of 
(1) a query step $s$ and 
(2) a query variable $v$ read by $s$,
\tool computes the data dependency of $s$ regarding $v$
as the last runtime definition of $v$ on $\tau$ before $s$.
In this work, we allow the user to specify implicit query variable  (e.g., the library variable used in \texttt{list.get(i)}).

While LLMs have demonstrated great potential,
they also renders randomness and hallucination.
Thus, to compute a dynamic data dependency based on trace, 
our challenges lie in that
(1) how to \textit{precisely} recover the unrecorded (implicit) variables regarding both their structure and values; and
(2) how to align the memory address of recorded and recovered variables for definition analysis.
%To this end,
%\tool decomposes the subject program $P$ into an instrumented part $P_i$ (e.g., application code) and
%an uninstrumented part $P_u$ (e.g., external library or remote service) where
%(1) $P_i$ can call APIs or functions defined in $P_u$ and
%(2) the execution of $P_i$ is recorded as the \textit{partial} trace $\tau$ with explicit variables' value and memory address recorded.
%Based on the slicing criteria,
%\tool recovers both miss recorded runtime variables and execution in $P_u$
%for computing dynamic data dependency.
%for constructing target data dependency.
%To this end, we design \tool in a two-stage manner,
%i.e., variable recovery stage and definition inference stage.
%As for recovering the variable recovery stage,
%if the query variable is implicit,
More specifically, to have precise dependencies, 
we need to recover both the value and the structure of the query variable in the query step.
For example, for 
a query step on the code of ``\texttt{map.getSize()}'' (e.g., with \texttt{map} of recorded value ``\{\texttt{key1=1, key2=2}\}'' and memory address of \texttt{x001}) and
an implicit query variable \texttt{size} inside \texttt{map},
%and an implicit query variable \texttt{size} which is internally defined in the \texttt{map},
\tool needs to infer
(1) the value of the query variable, \texttt{size}, as \texttt{2} and
(2) the its structure regarding \texttt{map} (i.e., \texttt{map.table.size}).

%The relation allows \tool to link the recovered variable to the variables with recorded memory address,
%mitigating potential alias inference challenges.
To enforce LLMs to have precise inference on the trace,
we propose an adaptive context generation technique
for LLM to infer code dynamics in different code scenarios.
Observing that the input context plays a crucial role for LLMs' prediction,
we design \tool to generate the most relevant execution examples at runtime for LLMs to follow.
Specifically, before inferring the runtime dynamics (e.g., values and structures of a query variable) with LLM, 
we synthesize a tiny program from the target code, with very small cost to be fully instrumented.
Then, recorded code dynamics can serve as the contextual examples for LLM to follow to infer the code dynamics.
By this means, 
\tool can have reliable predictions without any manually defined library summary or knowledge.

% a prompt
% synthesized from the code of the query step,
% to informatively indicate how to expand the query variable.
% In the definition inference stage,
% \tool
% (1) narrows down all the candidate steps potentially defining the query variable
% based on its relevant memory address and
% (2) use LLM to extrapolate the execution of the unrecorded call
% to confirm the true definition.
% By this means, \tool can
% compute the dynamic data dependency on the partial execution of a program,
% without the need of re-execution and re-instrumentation.
We extensively evaluate \tool against the state-of-the-art slicers such as Slicer4J, ND-Slicer, LLM Slicer, and re-execution Slicer on a total number of 8300 data-dependencies over 3 slicing benchmarks.
The results show that \tool can significantly outperform the baselines. The accuracy and recall, achieving 80.3\%, 91.1\%, and 98.3\% on the three benchmarks, whereas the best baseline reaches 39.0\%, 82.0\%, and 59.9\% (accuracy), and 53.4\%, 79.1\%, and 87.1\% (recall), respectively.
In addition, the performance is stable in different foundation models (ChatGPT-4o and Gemma-3).
We integrate \tool in a dual-slicing based approach for localizing regression bugs,
significantly improving its location performance by finding 16\% more regressions.

We summarize our contribution as follows:
\begin{itemize}[leftmargin=*]
  \item {\textbf{Methodology.}}
    To this best of our knowledge, we propose, \tool, 
    the first LLM-based dynamic data dependency analyzer on partial program trace. 
    Comparing to the state-of-the-art program dependency analyzer \cite{ahmed2021slicer4j} which infers the dependencies from the static code,
    we use LLM to recover the code dynamics from both source code and the partial trace for higher accuracy.
    In addition, we propose adaptive context generation technique to minimize the hallucination effect of LLMs on program analysis.
    %\tool adopts an LLM-based two-stage solution to infer dynamic data dependency
    %without the needs of re-execution and re-instrumentation.
  % \item \textbf{Dataset.}
  %   We deliver the largest dataset of dynamic data dependencies.
  %   Comparing to the existing datasets, our dataset is fine-grained into \textit{step} and concrete variable,
  %   consisting of \linyun{XX} data dependencies over \linyun{XX} distinct APIs from \linyun{XX} projects.
  %   The dataset can facilitate new advance in data dependency analysis in the community.
  \item {\textbf{Tool.}}
    Based on our technique, we deliver \tool as an open-source tool,
    which is designed to 
    (1) support both online and offline foundation models such as ChatGPTs and Gemma and
    (2) serve as a plugin for time-travelling debuggers \cite{wang2019explaining, sumner2013comparative, johnson2011differential, weeratunge2010analyzing, meinicke2018understanding}. 
    The source code, demo, and tutorials are available at \cite{recov-slicing}.
  \item {\textbf{Evaluation.}}
    We extensively evaluate \tool against the baselines solutions such as Slice4J, ND-Slider, and LLM-Slicer on the three benchmarks, showing that \tool can serve as a new state-of-the-art. 
    Compared to those ML-based slicing tools, 
    \tool enjoys its advantage especially on the dependencies over the unseen library calls.
    In addition, we integrate \tool with a dual-slicing based regression localizer,
    significantly improving its performance of locating regressions.
\end{itemize}

Given the space limit, more details (tool demo, source code, and experimental details) are available at \cite{recov-slicing}.

\section{Motivating Example}\label{sec:motivating-example}

\begin{figure*}[t]
  \centering
\begin{lstlisting}[language=Java]
void main(String[] args) {
  List<StringBuilder> sharedList = new ArrayList<>();
  StringBuilder originalRef = new StringBuilder("0");
  sharedList.add(originalRef); //sharedList.toString() = [''0'']

  StringBuilder aliasRef = sharedList.get(0); //sharedList.toString() is [''0'']
  if(Instant.now().getEpochSecond()%8 == 0) 
    aliasRef.append("0"); //sharedList.toString() is [''00'']

  if(Instant.now().getEpochSecond()%16 == 0)
    appendToAll(sharedList, "2"); //sharedList.toString() is [''002'']

  Assert(sharedList.get(0).toString(), "012"); //the actual value is "002", leading the query variable to be second character. 
}

void appendToAll(List<StringBuilder> list, String suffix) {
  for (StringBuilder item : list)
    item.append(suffix);
} 
\end{lstlisting}
  \caption{A simplified code example for computing dynamic data dependency.
  The query step is on line 13 and query variable is \texttt{sharedList.get(0).charAt(1)},
  its dynamic data dependency is the step on line 8.
  The comment indicates the values of the variables after executing the code.
  }\label{fig:motivating-example-code}
\end{figure*}

%\lstinputlisting[language=Java, caption={Example}, label=lst:motivating-example-code]{code/motivating-example-code.tex}

\autoref{fig:motivating-example-code} shows a simplified Java code example from a Defects4J bug \cite{just2014defects4j},
where the execution information is showed as comment in each line.
In the example, \texttt{shared\-List.\-get(0).\-char\-At(1)} should be `1' instead of '0' (at line 13).
We can reach the root cause (i.e., line 8)
if we can compute the dynamic data dependency of the query step on line 15
regarding the query variable \texttt{shared\-List.get(0).\-char\-At(1)}.
The dynamics of one run of the code in \autoref{fig:motivating-example-code} is showed in \autoref{fig:motivating-example-graph}.
The problem is challenging as follows:
\begin{itemize}[leftmargin=*]
  \item \textbf{Instrumentation Cost.}
    Probing the value and the structure of the query variable, \texttt{shared\-List.\-get(0).\-char\-At(1)},
    requires to instrument the library class \texttt{java.\-lang.Array\-List} and
    \texttt{java.\-lang.\-Str\-ing\-Builder},
    both are widely used primitive JDK classes.
    Their instrumentation in practice can introduce significant runtime overhead and risk of bytecode manipulation,
    broadly impacting irrelevant parts of the program.
  % \item \textbf{Alias Inference Challenge.}
  %   Even if we want to \textit{conditionally} instrument the query variable and its relevant variables,
  %   we need to infer the alias relation among the variables such as \texttt{originalRef} (defined in line 5), \texttt{aliasRef} (defined in line 6), and \texttt{sharedList.get(0)} (see purple ellipses in \autoref{fig:motivating-example-graph}).
  %   Inferring the alias upon assignment and object access, regarding potential polymorphism, is non-trivial.
  \item \textbf{Replication Feasibility/Cost.}
    The conditions in line 7 and line 10 depend on the timestamp to run the program,
    which means that
    we may not observe the bug in the next run.
    Re-executing the program can hardly guarantee that the interested \textit{dynamic} data dependency can be replicated.
\end{itemize}
The above challenges are prevalent among modern programming languages,
where their programs are built upon enriched external libraries with object variables refer to each other.
% In addition, the runtime dynamics may not always be deterministic.

Those challenges also largely limit the performance of the state-of-the-arts solutions such as Slice4J \cite{ahmed2021slicer4j}, ND-Slicer \cite{yadavally2024predictive}, and LLM-Slicer \cite{shahandashti2024program}.
As for Slice4J,
it can instrument the application code, i.e., steps and variables in the solid rectangle in \autoref{fig:motivating-example-graph}.
In \autoref{fig:motivating-example-graph}, we can see that only
variables \textit{explicitly} used in the application code are recorded,
which is \textit{not} informative to infer our expected dynamic data dependency.
In this example, it is the set of \textit{implicit} variables referenced by those explicit variables
carry more causalities among the steps.
Moreover, it is not realisitic to manually summarize all the library calls,
especially for those less popular libraries.
%In practice, many interested query variables are implicit and defined in the library.
As for ND-Slicer and LLM-Slicer,
they do not run the program for computing the data dependency.
Instead, they rely on their trained or adopted LLM to parse the program runtime behaviors based on its source code.
Those approaches are limited when the target programs have complicated method calls and loops (e.g., line 11 and line 16-18 in \autoref{fig:motivating-example-code}) and alias problem (e.g., the variables of \texttt{originalRef}, \texttt{aliasRef}, and sharedList.get(0)).
More details of running those tools on the example are available at \cite{recov-slicing}.
%As a result, they return a false positive result of line 2 in this case.

%In this work, we propose \tool to address the above challenges.
Different from ND-Slicer and LLM-Slicer,
\tool records light-weighted application code (see solid rectangle in \autoref{fig:motivating-example-code}),
as contextual information for LLM to simulate the execution more accurately.
Different from Slice4J,
\tool recovers miss-recorded variables (see dashed rectangle in \autoref{fig:motivating-example-code}) based on the recorded context and how they are \textit{indirectly} defined and used by the recorded steps with LLM.
%While the idea is intuitive,
The technique is non-trivial in two folds:

\begin{itemize}[leftmargin=*]
  \item \textbf{Variable Structure Complication.}
    The runtime variable is structural (e.g., array, map, set, etc.),
    which can have dynamic depth and breadth,
    LLM-based solution needs to overcome its inherent hallucination to
    (1) \textit{accurately} recover the \textit{object graph} in the execution and
    (2) identify the graph node matching the query variable.
  \item \textbf{Linking the Recorded and the Recovered Variables.}
    The recovered variable can share the same memory address with the recorded variables
    (see the dashed purple ellipse and the solid purple ellipse).
    Therefore, \tool needs to link the recovered and the recorded to
    facilitate variable definition inference.
\end{itemize}

% As for the former, we design a novel adaptive in-context learning approach to enforce LLM to
% generate useful object graph by feeding LLM with a synthesized prompt relevant to the query step.
% As for the latter, we narrow down the candidate steps relevant to the query variable
% and use LLM to simulate the miss-recorded execution to infer variable alias and definition.
In this \tool work, we guide the LLM with adaptive prompts on runtime execution to build the relevant object graph, by simulating missing execution steps, recover a query variable’s aliases and definitions.
More details are illustrated in Section~\ref{sec:approach}.

\section{Preliminaries and Problem Statement}

\begin{figure}[t]
  \centering
  \includegraphics[width=1.0\linewidth]{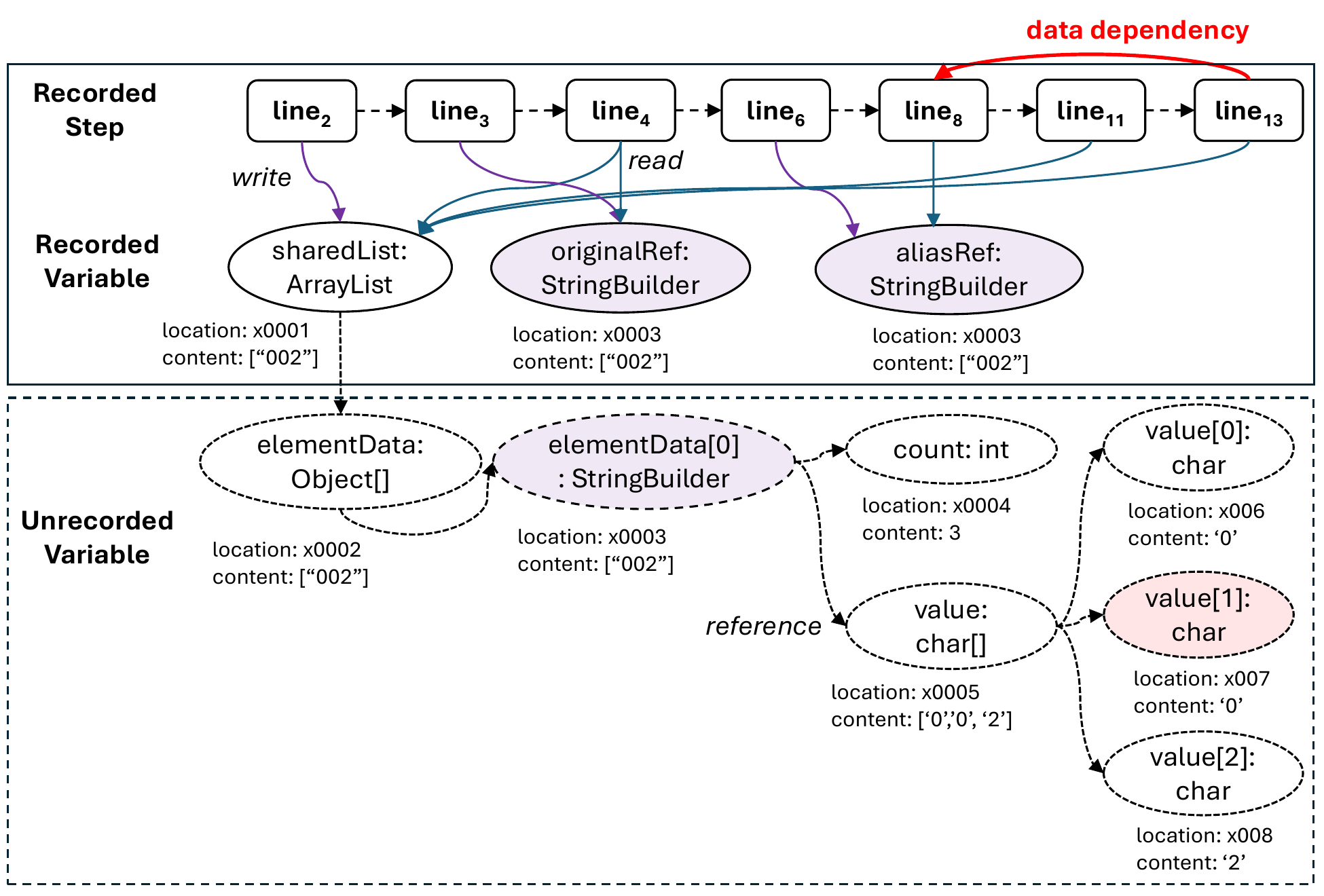}
  \caption{The dynamics of the code in \autoref{fig:motivating-example-code}, assuming that only the \texttt{main()} method is recorded.
  The dynamic data dependency (in red) between the step on line 8 and the step on line 13 is non-trivial to compute.
  The purple lines indicate \textit{define} relations, the blue lines indicate \textit{use} relations, the solid ellipses indicate the recorded variables, the dashed ellipses indicate the unrecorded variables, and the dashed lines indicate \textit{reference} relations.
  The purple ellipses are the variables sharing the same memory address; and
  the red ellipse is the query variable to be recovered.
  }\label{fig:motivating-example-graph}
\end{figure}

\noindent\textbf{Program.}
Given a program $\mathcal{P} = \langle \mathcal{N}, \mathcal{V}, \mathcal{E}_{flow}, \mathcal{E}_{read}, \mathcal{E}_{written},\\
\mathcal{E}_{ref} \rangle$
where $ \mathcal{N}$ is the set of instructions;
$\mathcal{V}$ is the set of variables (of name and type),
$\mathcal{E}_{flow} \subset \mathcal N \times \mathcal N$ is the set of possible execution flows from one instruction $n_i$ to another instruction $n_j$ ($n_i, n_j\in \mathcal{N}$),
$\mathcal{E}_{read} \subset \mathcal{N}\times \mathcal{V}$ is the set of relations where an instruction $n\in \mathcal{N}$ reads a variable $v\in \mathcal{V}$;
$\mathcal{E}_{written} \subset \mathcal{N}\times \mathcal{V}$ is the set of relations where an instruction $n\in \mathcal{N}$ writes a variable $v\in \mathcal{V}$; and
$\mathcal{E}_{ref} \subset \mathcal{V}\times \mathcal{V}$ is the set of reference relations where a variable $v_i\in \mathcal{V}$ refers to another variable $v_j\in \mathcal{V}$.

\noindent\textbf{Execution.}
Given a program $\mathcal{P}$ and one input $\mathcal{I}$,
% based on a predefined operation semantics\footnote{In this work, we follow the common operation semantics of modern language design.},
we use $\mathcal{T} = (\tau, \mathcal{V}al)$ to denote 
the \textbf{trace model}, i.e., full execution of $\mathcal{P}$ under $\mathcal{I}$.
Specifically, we define a trace $\tau = \langle s_1, s_2, ..., s_k \rangle$ where:
\begin{itemize}[leftmargin=*]
  \item \textbf{Step.} $s_i\in \tau$ is an occurrence of an instruction $n_i\in \mathcal{N}$, which can be further denoted as $s_i = (n_i, order)$ where $order\in \mathbf{Z}^+$ is the occurrence order of $n_i$;
  \item \textbf{Flow Instance.} $(s_i, s_j)$, $s_i, s_j\in \tau$, is an occurrence of a flow $e\in \mathcal{E}_{flow}$;
\end{itemize}
We define a set of variable instances $\mathcal{V}al = \{val_1, val_2, ...,\\ val_h\}$  where:
\begin{itemize}[leftmargin=*]
  \item \textbf{Variable Instance.} $val_i\in \mathcal{V}al$ is an instance of a variable $v\in \mathcal{V}$,
    denoted as $val_i = (v_i, content, m)$ where 
    $content$ is the concrete value of $v$ and
    $m$ is the memory location of $v$.
    In addition, we assume that any value $content$ can be presented in a string form, e.g., ``0'', ``hello world'', and ``[1, 2, 3]''.
    
  \item \textbf{Read/Write Instance.} A relation instance $(s_i, val_j)$, $s_i\in \tau$, $val_j\in \mathcal{V}al$, is an occurrence of a relation $r\in \mathcal{E}_{read} \cup \mathcal{E}_{written}$;
  \item \textbf{Reference Instance.} A relation instance $(val_i, val_j)$, $val_i, val_j\in \mathcal{V}al$, is an occurrence of a relation $r\in \mathcal{E}_{ref}$.
      If a variable instance $val_i$ refers to another variable instance $val_j$, its content can recursively contain the value of its reference. 
      %we let the content of $val_i$ refer to the content of $val_j$.
      For example, if an array variable instance refer to three array elements of content \texttt{"1", "2"}, and \texttt{"3"}, its content can be \texttt{["1", "2", "3"]}.
      \footnote{The variable content can be achieved by implementations of the \texttt{toString()} in Java, \texttt{\_\_str\_\_()} in Python, and \texttt{std::to\_string()} in C++.}.
\end{itemize}

%, \texttt{to\_string()} in Rust, and \texttt{std::to\_string()} in C++.

For the example in \autoref{fig:motivating-example-graph}, 
we use a rounded rectangle to represent a step,
which can read and write a variable instance (in ellipse).
A variable instance \texttt{sharedList} refers to a variable instance \texttt{elementData} of an array type,
which in turn refers to its element \texttt{elementData[0]} of a \texttt{StringBuilder} type.
In addition, the variable \texttt{originalRef}, \texttt{aliasRef}, and \texttt{elementData[0]} share
the same memory location, i.e., they are alias variables.

\noindent\textbf{Dynamic Data Dependency.}
Given two steps $s_i$, $s_j \in \tau$, we say that $s_i$ is dynamically data-dependent on $s_j$ regarding a variable instance $val$ if $s_i$ reads a variable instance $val$ defined by $s_j$, and $\nexists$ $s_k$ between $s_j$ and $s_i$ that defines any variable instance $val_i$ sharing the same memory location with $val$.

Obviously, if we can \textit{ideally} have a complete trace model $\mathcal{T} = (\tau, \mathcal{V}al)$,
it is straightforward to compute any dynamic data dependency.
However, acquiring a full trace model is prohibitively costly.
Therefore, we usually have \textit{partial} trace model (as in the solid rectangle in \autoref{fig:motivating-example-graph}),
providing insufficient information to compute the dynamic dependency.
%Next, we provide a more realistic settings for our problem.

%\noindent\textbf{Dynamic Control Dependency.}
%Given two steps $s_i$, $s_j \in \tau$, $s_i$ is control-dependent on $s_j$ if $s_j$ controls the execution of $s_i$, and $\nexists$ $s_k$ between $s_j$ and $s_i$ that controls the execution of $s_i$.

\noindent\textbf{Program Partition.}
Assuming that a program $\mathcal{P}$ can be partitioned into two parts, i.e., 
the instrumented part $\mathcal{P}_i$ and
the uninstrumented part $\mathcal{P}_u$, denoted as
$\mathcal{P} = (\mathcal{P}_i, \mathcal{P}_u, \mathcal{E}_{call}, \mathcal{E}^*_{ref})$, where
$\mathcal{P}_i = \langle \mathcal{N}_i, \mathcal{V}_i, \mathcal{E}_{flow_i}, \mathcal{E}_{read_i}, \mathcal{E}_{written_i}, \mathcal{E}_{ref_i}\rangle$,
$\mathcal{P}_u = \langle \mathcal{N}_u, \mathcal{V}_u, \mathcal{E}_{flow_u},\\ \mathcal{E}_{read_u}, \mathcal{E}_{written_u}, \mathcal{E}_{ref_u}\rangle$,
%$\mathcal{P}_i = (\mathcal{N}_i, \mathcal{V}_i)$, $\mathcal{P}_u = (\mathcal{N}_u, \mathcal{V}_u)$,
$\mathcal{E}_{call} \subset \mathcal{N}_i \times \mathcal{N}_u$ and
$\mathcal{E}^*_{ref} \subset \mathcal{V}_i \times \mathcal{V}_u$.
$\mathcal{E}_{call}$
indicates that the instrumented partition $\mathcal{P}_i$ calls the uninstrumented partition $\mathcal{P}_u$.
Given $(n_i, n_u)\in \mathcal{E}_{call}$, we call $n_i$ is a \textbf{call site} in $\mathcal{P}_i$ to $\mathcal{P}_u$.
For example, line 8 in \autoref{fig:motivating-example-code} is a call site in the instrumented code (i.e., $\mathcal{P}_i$) to the library code \texttt{StringBuffer.append()} ($\mathcal{P}_u$). 
If a step $s_i\in \tau_i$ running on a call site leads to the execution of $s_u$ running on $n_u\in \mathcal{N}_u$,
we say $s_i$ is a \textbf{call site step} to $s_u$.

\noindent\textbf{Program Statement.}
Given an input $\mathcal{I}$ to $\mathcal{P}$ (with only $\mathcal{P}_i$ instrumented), we can have a partial trace model $\mathcal{T}_i = (\tau_i, \mathcal{V}al_i)$. %where
%it can have steps, variable instances, and relation instances on $\mathcal{P}_i$.
%Assume the availability of full trace model $\mathcal{T}^* = (\tau^*, \mathcal{V}al^*)$ of $\mathcal{P}$,
%we can define \textit{call instance} $(s_i, s_u)$ where
%(1) $s_i\in \tau_i$, $s_u\notin \tau_i$, and $s_u\in \tau^*$ and
%(2) $s_i$ is the closest step calling the instruction of $s_u$ before $s_u$.
%In brief, $s_i$ actually calls $s_u$ in the execution.
Given $\mathcal{P} = (\mathcal{P}_i, \mathcal{P}_u)$, the partial trace model $\mathcal{T}_i$,
a query step $s_q\in \tau_i$, a query variable instance $val_q$ ($val_q$ is not necessary in $\mathcal{V}al_i$),
we compute the dynamic data dependency of $s_q$ and $val_q$ as follows.

Assume that the ground-truth full trace model is $\mathcal{T}^* = (\tau^*, \mathcal{V}al^*)$ of $\mathcal{P}$,
we can have $s^* \in\tau^*$ which is dynamic data-dependent by $s_q$ regarding $val_q$:

\noindent\textbf{Case 1}: If $s^*\in \tau_i$ (i.e., $s^*$ is instrumented), we return $s^*$.

\noindent\textbf{Case 2}: If $s^*\notin \tau_i$ (i.e., $s^*$ is not instrumented), we return $s_i$ as the call site step of $s^*$.

% \begin{itemize}[leftmargin=*]
%   \item If $s^*\in \tau_i$ (i.e., $s^*$ is instrumented), we return $s^*$.
%   \item If $s^*\notin \tau_i$ (i.e., $s^*$ is not instrumented), we return $s_i$ as the call site step of $s^*$.
% \end{itemize}

For example in \autoref{fig:motivating-example-graph}, 
given the query step on line 13 and the query variable \texttt{shared\-List.\-element\-Data[0].value[1]}, 
we return the step on line 8, i.e., \texttt{aliasRef.append("0")}
which is the call site step to the library code to define the variable instance.
Note that, the ground-truth trace model $\mathcal{T}^*$ is not available in practice, 
the problem is generally on how to recover the minimal information in $\mathcal{T}^*$ based on $\mathcal{P}$ and $\mathcal{T}_i$ for computing the dependency.
\section{Approach}\label{sec:approach}

\begin{figure}[t]
  \centering
  \includegraphics[width=1.0\linewidth]{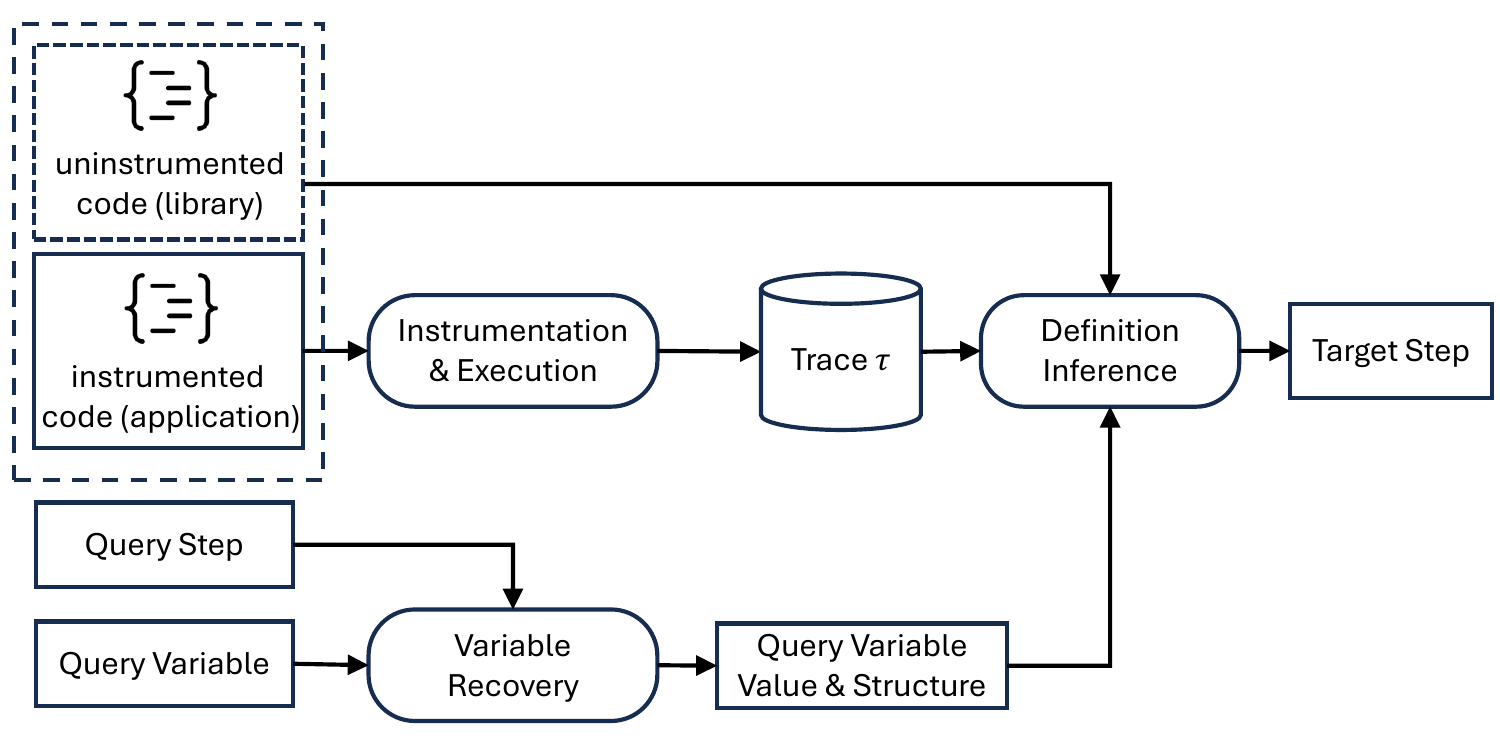}
  \caption{
  %\textcolor{red}{Comment 2: instrumented code -> execution (instead of instrumentation?) -> trace -> ...}
  The approach overview of \tool, takes input as a query step and a query variable, and
  generate their dynamic data dependency as output.
  An ellipse represents a process and a rectangle represents its input or output.
  %Starting with a focal step, a time-travelling agent can repetitively issue
  %queries (e.g., data or control slicing) on the trace steps toward the root cause.
  The approach consists of
  (1) variable recovery for the value and memory structure of the variable and
  (2) definition inference to narrow down the scope of candidate steps for LLM to confirm the true runtime definition.
  }
  \label{fig:overview}
\end{figure}

\noindent\textbf{Overview.}
\autoref{fig:overview} shows the overview of our approach \tool.
First, a subject program $\mathcal{P}$ is partitioned into instrumented and uninstrumented parts\footnote{In the implementation, we partition the program into application and library code as instrumented and uninstrumented parts by default.}
where the instrumented part can derive a partial trace $\tau$.
Given a user-specified query step $s_q$ on $\tau$ and its query variable instance $val_q$ (which can be an implicit variable in the uninstrumented part),
%\footnote{In this work, we allow the user to specify the query variable instance},
we design a \textbf{Variable Recovery} technique to recover its value as well as how it could refer or be referred to by
existing recorded variables.
Based on the recovered variable, 
we design a \textbf{Definition Inference} technique to identify a step on trace as its dynamic data dependency.

\subsection{Variable Recovery}\label{sec:variable-expansion}

\begin{figure}[t]                  % “t” = try top of page
  \centering
  \[
  \begin{array}{rcll}
    \langle\textit{Path}\rangle
      &::=& \langle\textit{Var}\rangle
      &\text{\small(root variable)}\\[2pt]
      &\mid& \langle\textit{Path}\rangle\;\texttt{.}\;\langle\textit{Field}\rangle
      &\text{\small(field access)}\\[2pt]
      &\mid& \langle\textit{Path}\rangle\;\texttt{[}\,\langle\textit{Expr}\rangle\,\texttt{]}
      &\text{\small(array or list index)}
  \end{array}
  \]
  \caption{The format of query variable instance, which is an access path from the root variable to target variable.}
  \label{fig:access-path-syntax}
\end{figure}

%\noindent\textbf{Query Variable Instance.}
As showed in line 13 in \autoref{fig:motivating-example-code},
where the query variable instance is the second character of the \texttt{StringBuilder} object inside the variable \texttt{sharedList} of \texttt{ArrayList} type,
we recover the query step with its value (i.e., \texttt{'0'}) and structure (i.e., \texttt{shared\-List.element\-Data[0].val\-ue[1]}) from the query step regarding its recorded variable \texttt{sharedList} and value \texttt{["002"]}.
\autoref{fig:access-path-syntax} shows the syntax to specify a query variable instance in our work.
Generally, we aim to recover from the query step the \textit{object graph}
where
a variable instance (e.g., \texttt{sharedList}) is its root and
the query variable instance is one of its nodes.
Formally, given root variable instance $val_1$, we would like to recovery the variable instances in $\mathcal{E}_{ref}^* (\{val_1\})$ to construct a path from $val_r$ to variable instance $val_q$, where $val_q$ is the actual query variable instance\footnote{Here, $\mathcal{E}_{ref}^*$ is the closure relation of $\mathcal{E}_{ref}$ and $\mathcal{E}_{ref}^* ( \cdot )$ is to apply the relation to a set.}.
\autoref{fig:motivating-example-graph} shows an example of object graph structure. %Our method will recover the path from \texttt{sharedList} to \texttt{value[1]}.

\noindent\textbf{Challenge.}
Intuitively, LLMs (e.g., ChatGPTs) can parse variable instance and query step as a sequence of tokens and
generate the expected object graph (e.g., in a format like JSON).
However, when the to-be-recovered object graph has a large depth and breadth,
the recovering performance is limited.
As our observation, important variable content could be missed and mis-modified.
In addition, the generated token sequence representing object graph can sometimes be ill-formatted.
Given the space limit, audience can refer to \cite{recov-slicing} for more examples on native prompt engineering.

\begin{figure}[t]
  \centering
  \includegraphics[width=1.0\linewidth]{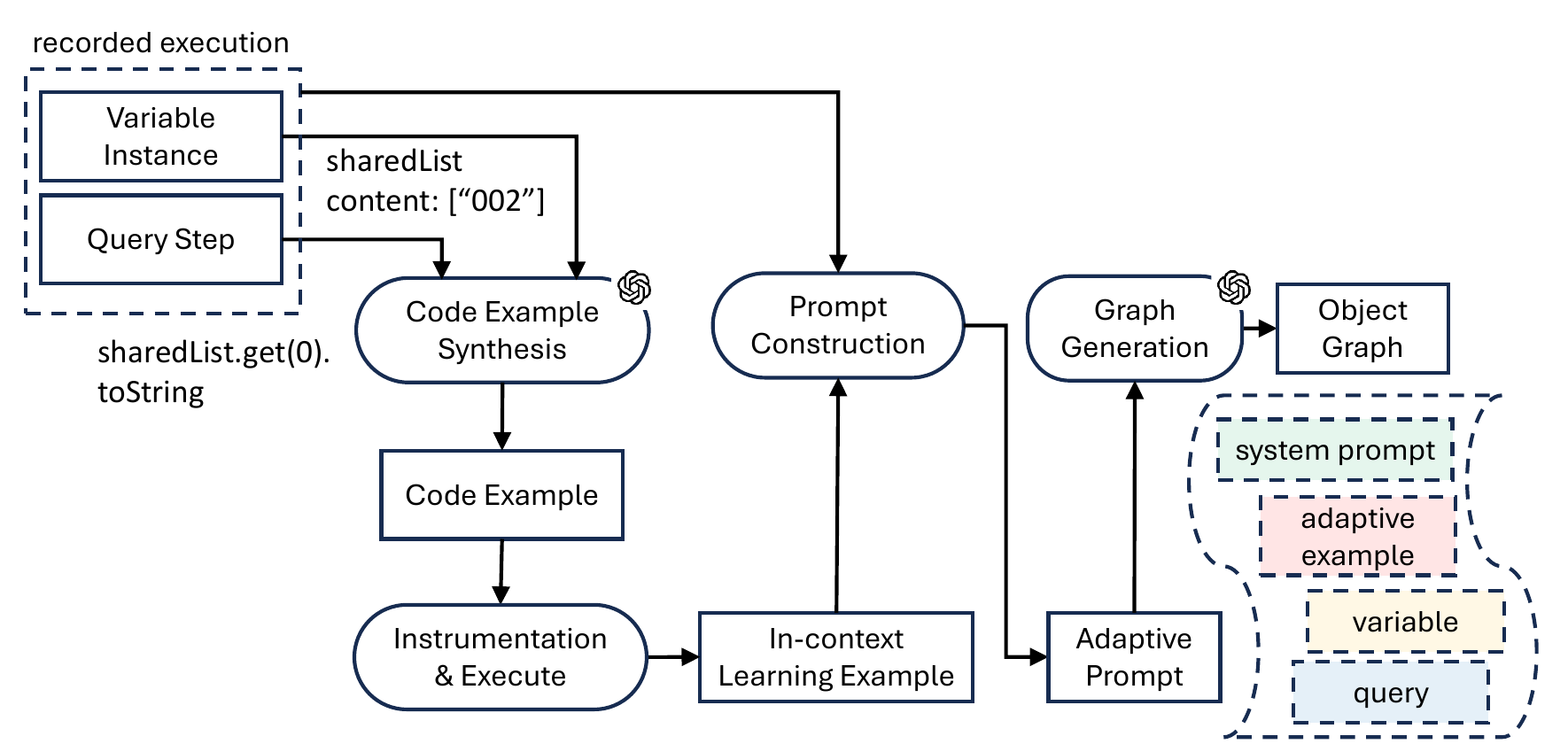}
  \caption{Overview of variable recovery with adaptive in-context learning technique:
  We construct an adaptive prompt based on the query step and construct an object graph whose root is the given variable instance.
  }\label{fig:variable-expansion}
\end{figure}

\begin{sloppypar}
\noindent\textbf{Adaptive Context Generation Solution.}
We design an adaptive context generation to facilitate in-context learning \cite{dong2022survey} to recover the variable value and structure.
%Specifically, given a LLM as $f(.)$ and its question $q$, NLP researchers have observed that
%if we can find a relevant example $p$ as an additional context to $q$,
%$f(p:q)$ could be significantly more accurate than $f(q)$.
%Compared to the traditional in-context learning solution which the context $p$ is pre-defined,
%we propose an adaptive in-context learning approach where
%we synthesize the context $p$ based on the variable instance and the query step.
\autoref{fig:variable-expansion} shows an overview of our solution.
Given the code of a query step, e.g., \texttt{sharedList.get(0).\\toString()}, and the string content of variable instance, e.g., \texttt{["002"]},
we first generate a small executable code $\mathcal{P}'$ ended with the code of query step so that $P'$ access a similar variable instance $val'$.
$\mathcal{P}'$ is controlled within $t$ lines of code, where $t$ is a user-defined threshold.
Then, we apply full instrumentation on the generated $\mathcal{P}'$ 
so that we can have the object graph of $val'$ as the in-context learning example.
\autoref{fig:prompt-example-code} shows a synthesized code example,
whose full instrumentation into relevant library code allows us to have a similar object graph $G'$ with $val'$ as its root.
Then, we construct the adaptive example consisting of 
(1) an input field as variable instance $val'$ and 
(2) an output field as object graph $G'$.
Finally, we construct the graph-generation prompt consisting of a system prompt, adaptive example, variable instance, and query for LLM to generate a relevant object graph.
% Some data structures may have complex internal representations that are difficult to model accurately. For instance, in Java, a \texttt{HashMap} is internally represented as a list of hash slots, each containing detailed key-value pairs, making it challenging for LLMs to predict all details. To address this, during variable recovery, we prompt LLMs to recover only the relevant variables along the access path to the target $val_q$, and we use an abstract representation for the given variable.
Given the space limit, the concrete prompts are available in appendix \ref{sec:prompt-varrecovery}.

\end{sloppypar}

\begin{figure}[t]
  \centering
\begin{lstlisting}[language=Java]
void main(String[] args) {
  List<StringBuilder> sharedList = new ArrayList<>();
  sharedList.add(new StringBuilder("002"));
  Assert(sharedList.get(0).toString(), "012"); 
}
\end{lstlisting}
  \caption{A synthesized code example for adaptive example, based on the code of query step and the content of variable instance.
  Full instrumentation into line 4 allows us to have an object graph similar to the one in \autoref{fig:motivating-example-graph}.
  }\label{fig:prompt-example-code}
\end{figure}

\subsection{Definition Inference}
Given a recovered query variable instance $val_q$ and how a recorded variable instance (as root) $root$ can access $val_q$, denoted as $root.val_1.\\value_2...val_q$,
we find the definition step of $val_q$ in the recorded trace.
In this section, we illustrate our approaches with formal algorithms in the paper.
For more readability, the animations of the algorithms are available in \cite{recov-slicing}.

\noindent\textbf{Challenge.}
Note that, LLMs can only recover value of variable instance, instead of their memory location.
Therefore, for a \textit{reference path} such as $root.val_1.value_2...val_q$,
only the variable instance $root$ has its memory location available.
Moreover, despite the memory address of intermediate variable instance on the reference is not known,
it could be alias to some explicit (i.e., recorded) variable instance in the partial trace $\tau$.
Any steps manipulating an intermediate variable instance or its alias variables (e.g., by calling a function defined in uninstrumented part $\mathcal{P}_u$)
can be a potential candidate step for defining $val_q$.

%%% Overall Algorithm
\begin{algorithm}[tbp]
\caption{Overall Definition Step Search Algorithm}
\label{alg:overall}
\begin{algorithmic}[1]

\STATE \textbf{Inputs:} 
    a query step $s_q$, 
    a reference path $p = \langle root, val_1, val_2, ..., val_q \rangle$, and
    a partial trace $\tau$
\STATE \textbf{Output:} the definition step $s_{def}$ for the query step $s_q$ and query variable instance $val_q$

{\color{ForestGreen} // alias inference}
\STATE $aliasMap$ := $\emptyset$ 
\FOR {$val\in p$ from root to leaf}
    \STATE memory location $loc$, $V=\{val_{a1}, val_{a2}, ...\}$ := \textbf{inferAlias}($aliasMap$, $val$, $p$, $\tau$)
    \STATE $aliasMap$.put($s$, $V$)
\ENDFOR

{\color{ForestGreen} // search for definition step}    
\FOR {$s\in \tau$ in a backward manner}
    \IF {\textbf{isDef}($aliasMap$, $val_q$, $p$, $s$) == true}
        \RETURN $s$
    \ENDIF
\ENDFOR

\RETURN null

\end{algorithmic}
\end{algorithm}

%\begin{algorithm}
%\caption{Recov-Slicing Algorithm}
%\label{alg:overall}
%\begin{algorithmic}[1]
%
%\STATE \textbf{Inputs:} $step_f$, $var_f$, $S_v$
%\STATE \textbf{Output:} $step_{def}$
%
%\STATE $step_{def}$ := recorded defining step of $var_f$
%{\color{ForestGreen} // Data slicing} % comment
%
%% comment
%{\color{ForestGreen} // Not deadend $\Rightarrow$ continue slicing}
%
%\IF {$var_f.value$ in $step_{def}$ is different from expectation}
%    \STATE return $step_{def}$
%\ENDIF
%
%% comment
%{\color{ForestGreen} // Deadend $\Rightarrow$ trace recovery (find $step_{def}$)}
%
%\STATE $S_{knownAddress}$ := \textbf{inferAliasRelations}($step_f$, $step_{def}$, $S_v$)
%\STATE $step_{def}$ := \textbf{inferDefiningRelations}($var_f$, $step_f$, $step_{def}$, $S_{knownAddress}$)
%
%\STATE return $step_{def}$
%
%\end{algorithmic}
%\end{algorithm}

\noindent\textbf{Solution.}
To this end, we design Algorithm~\autoref{alg:overall} which consists of two steps,
i.e., 
(1) \textit{alias inference} for all the recorded variables sharing the same memory location with the variable instances on the reference path (line 3-7 in  Algorithm~\autoref{alg:overall});
(2) \textit{definition inference} for the true step dynamically defining the query variable instance (line 8-13).
In Algorithm~\autoref{alg:overall},
we maintain an alias map with the key as memory location and 
the value as the set of variable instances sharing the same location.
Thus, the alias map maintains all the variables instances 
alias with a variable instance in the reference path $p$,
Note that, any steps calling external code with variable instance in the alias map can potentially define $val_q$.
Thus, we check the definition relation for each candidate step in a backward manner (line 9).
% Note that, we search for the candidates  (line 8 in Algorithm~\ref{alg:overall}).

\begin{figure*}
    
  \centering
  \includegraphics[width=1.0\linewidth]{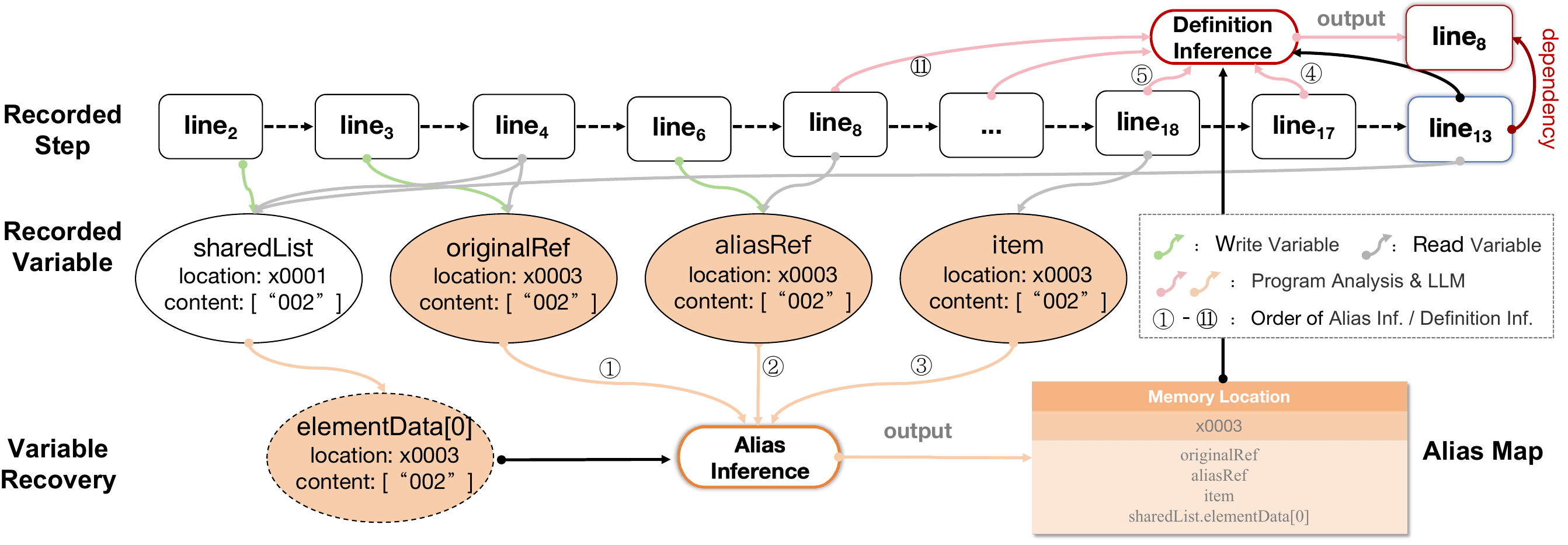}
  \caption{
  Run the motivating example using the algorithm. The query step is on line 13, and the query variable is sharedList.get(0).charAt(1). After running the three algorithms, we can find its dynamic data dependency step on line 8.
  }
  \label{fig:algorithm-motivating}
\end{figure*}

For example, consider the motivating case in \autoref{fig:motivating-example-code}. 
\autoref{fig:algorithm-motivating} illustrates the proposed algorithm step by step. 
Given the slicing criterion at $\mathbf{line_{13}}$, with \texttt{sharedList.elementData[0].value[1]} as the query variable instance, 
Algorithm~\ref{alg:overall} first scans the execution trace forward to construct the alias map (via Algorithm~\ref{alg:alias}), 
and then traverses the trace backward to locate the most recent definition step (via Algorithm~\ref{alg:def}). 
In this example, the algorithm ultimately identifies $\mathbf{line_{8}}$ as the defining step, 
since it modifies the target variable. 
This demonstrates how LLM-guided variable recovery and alias detection 
bridge the gap between partial traces and inferred execution semantics. 
In the following two subsections, we present the detailed implementations of 
alias inference and definition inference.

% \begin{figure}[t]
%   \centering
%   \includegraphics[width=0.6\linewidth]{acmart-primary-5/samples/image/algorithm 1.pdf}
%   \caption{
%   The motivating example runs on algorithm 1.
%   }
%   \label{fig:algorithm1}
% \end{figure}

% \input{definition-inference-algo.tex}

\noindent\textbf{Alias Inference.}
Algorithm~\ref{alg:alias} shows how to infer alias of a variable instance in the trace, combining traditional must-alias analysis and LLM-based inference.
We first adopt a light-weighted must-alias analysis for simple scenarios (line 6-7 in Algorithm~\ref{alg:alias}).
For more complex scenarios as a step $s_i$ calls external code (line 8),
we extract the external code and variable instance into a prompt for LLM to deduce the potential alias relation.
We design the prompt in a conservative way to avoid false positive (see appendix \ref{sec:prompt-alias}).

This subroutine detects all aliases of a target variable by scanning the trace and combining static must-alias checks with LLM inference for external code calls.

\newcommand*\circled[1]{\tikz[baseline=(char.base)]{
            \node[shape=circle,draw,inner sep=1pt] (char) {#1};}}

In our motivating example, as illustrated in \autoref{fig:algorithm-motivating}, 
the recovered variables are first passed to the \textbf{Alias Inference} module to identify potential alias relationships. 
This is achieved by iterating forward through each recorded step and querying the LLM 
to determine whether the recovered variables share alias relationships with the variables appearing in that step. 
In the example, \circled{1}, \circled{2}, and \circled{3} represent three alias inference operations, 
where the variables are predicted to be aliases of the recovered \texttt{elementData[0]}. 
These alias relationships are then inserted into the alias map, 
which serves as a crucial foundation for the subsequent definition inference phase.

%%% Alias Inference
\begin{algorithm}[tbp]
\caption{inferAlias Algorithm}
\label{alg:alias}
\begin{algorithmic}[1]

\STATE \textbf{Inputs:} 
    a target variable instance $val$, 
    a reference path $p$,
    an alias map $aliasMap$, 
    a trace $\tau$
\STATE \textbf{Output:} 
    a memory location $loc$,
    a set of variable instances with $loc$, $V$

\STATE $V$ := $\emptyset$

\FOR{$s_i \in \tau$}
    \STATE $val_{def}$ := $s_i$'s definition variable
    \IF{isMustAlias($s_i$, $val$)}
        \STATE $aliasMap$.put($val.loc$, $aliasMap$.get($val.loc$)$\cup$ \{$val_{def}$\})
    \ELSIF {$s_i$ is a call site and $s_i$ reads $val$, its parents in $p$, or their alias}
        \IF {isAliasWithLLMInference($s_i$, $val$)}
            \STATE $aliasMap$.put($val.loc$, $aliasMap$.get($val.loc$)$\cup$ \{$val_{def}$\})
        \ENDIF
    \ENDIF
\ENDFOR

\RETURN $val.loc$, $aliasMap.get(val.loc)$

\end{algorithmic}
\end{algorithm}

%\begin{algorithm}
%\caption{inferAliasRelations}
%\label{alg:alias}
%\begin{algorithmic}[1]
%
%\STATE \textbf{Inputs:} $step_f$, $step_{def}$, $S_v$
%\STATE \textbf{Output:} $S_{knownAddress}$
%
%\STATE $S_{knownAddress}$ := \{$root_f$\}
%
%\FOR{$step_i$ in ($step_{def}$ ... $step_f$)}
%    \STATE $S_{read_i}$ := set of variables read by $step_i$
%
%    % comment
%    {\color{ForestGreen} // Step to check: reads a variable with known address}
%
%    \IF{$\exists v \in S_{knownAddress}$ such that $v \in S_{read_i}$}
%        \STATE $S_{knownAddress}$ := $S_{knownAddress} \cup \{v|v \in S_{read_i} \cap v \text{ is alias of a variable in } S_v\}$
%        {\color{ForestGreen} // Record new alias relations} % comment
%    \ENDIF
%\ENDFOR
%
%\STATE return $S_{knownAddress}$
%
%\end{algorithmic}
%\end{algorithm}

% \begin{figure}[t]
%   \centering
%   \includegraphics[width=1.0\linewidth]{acmart-primary-5/samples/image/algorithm 2.pdf}
%   \caption{
%   The motivating example runs on algorithm 2.
%   }
%   \label{fig:algorithm2}
% \end{figure}

\noindent\textbf{Definition Inference.}
Algorithm~\ref{alg:def} shows how we infer the candidate definition step for a given query variable instance $val_q$.
We combine light-weighted static analysis for predefined simple scenario (line 3-4) and
LLM-based deduction for dealing with external code invocation (line 9).
Specifically, Algorithm~\ref{alg:def} first check whether it is a trivial case when 
the query variable is recorded and equal to the defined variable of the candidate step.
If not, we enumerate all the variable instances potentially can serve as a \textit{handle}
to define $val_q$ via calling external code (line 6-7).
 Similar to the prompt design in alias inference,
we conservatively design the prompt including the query variable, its reference path, and the external code (see appendix \ref{sec:prompt-def}).

In our motivating example, as shown in \autoref{fig:algorithm-motivating}, 
the \textbf{Definition Inference} phase is carried out by traversing the execution trace backward 
and examining each step to determine whether it defines the target variable. 
Specifically, steps \circled{4}, \circled{5}, $\cdots$, \circled{11} correspond to the backward search from $\mathbf{line_{13}}$ through $\mathbf{line_{12}}$, $\cdots$, down to $\mathbf{line_{8}}$. 
The algorithm ultimately identifies $\mathbf{line_{8}}$ (\texttt{aliasRef.append("0")}) as the defining step.

\begin{algorithm}[tbp]
\caption{isDef Algorithm}
\label{alg:def}
\begin{algorithmic}[1]

\STATE \textbf{Inputs:}
    an alias map $aliasMap$,
    a query variable instance $val_q$,
    a reference path $p$,
    a candidate step $s$
\STATE \textbf{Output:} a variable on whether $s$ defines $val_q$

\IF{$val_q$ is not recovered and $val_{def}.loc$ ==  $val_q.loc$}
    \RETURN true
\ELSIF {$s_i$ is a call site and $s_i$ reads a variable instance $val\in H$}
    \STATE $val_{def}$ := $s$'s definition variable
    \STATE $H$ := getHandle($p$, $aliasMap$)
    \STATE $call$ := the code of $s_i$'s call site
    \IF{any read variables of $s$ are in $H$ and isDefWithLLM($val$, $p$, $call$)}
        \RETURN true
    \ENDIF
\ENDIF

\RETURN false

\end{algorithmic}
\end{algorithm}

%\begin{algorithm}
%\caption{inferDefiningRelations}
%\label{alg:def}
%\begin{algorithmic}[1]
%
%\STATE \textbf{Inputs:} $var_f$, $step_f$, $step_{def}$, $S_{knownAddress}$
%\STATE \textbf{Output:} $step_{def}$
%
%% comment
%{\color{ForestGreen} // Candidate Steps: read a variable with known address}
%
%\STATE $S_c$ := $step$ in ($step_f$ ... $step_{def}$) such that $step$ reads $v \in S_{knownAddress}$
%
%\FOR{$step$ in $S_c$}
%    \IF{$step$ writes $var_f$}
%        \STATE $step_{def}$ := $step$
%        \STATE \textbf{break}
%    \ENDIF
%\ENDFOR
%
%\STATE return $step_{def}$
%
%\end{algorithmic}
%\end{algorithm}

% \begin{figure}[t]
%   \centering
%   \includegraphics[width=0.8\linewidth]{acmart-primary-5/samples/image/algorithm 3.pdf}
%   \caption{
%   The motivating example runs on algorithm 3.
%   }
%   \label{fig:algorithm3}
% \end{figure}

%\subsection{Validity of Approach}
%
%\hongshu{TODO: Identify properties that the baselines fail to hold, prove that \tool holds these properties}
%
%
%\subsection{Time Complexity}
%
%\subsection{Implementation}

\section{Evaluation}

% Research Questions
We evaluate the performance of \tool with the following research questions:

\noindent \textbf{RQ1. (Dynamic Slicing Performance)}
Is \tool effective in improving the performance of dynamic slicing compared to the state-of-the-art solutions?

\noindent \textbf{RQ2. (Component-Wise Analysis)}
How accurate are the predicted variable values and slices?

\noindent \textbf{RQ3. (Ablation Study)}
How does in-context learning of \tool contribute to the overall performance?

\noindent \textbf{RQ4. (Application Analysis)}
Does \tool improve the performance of typical application of dynamic slicing, such as Time-Travel Debugging?

% \hongshu{ Mention benchmark diversity \& metrics in later sections:
% Does \tool improve the dynamic slicing performance under the scenarios of large-scale programs with library calls, complex control flows, and non-deterministic programs? How much extra runtime overhead does it introduce? }

\subsection{Dynamic Slicing Performance (RQ1)}

%and RQ3, we constructed a diversity benchmark for evaluating the performance of
%\tool and the baselines.

\subsubsection{Setup}
To evaluate RQ1 and RQ3, we assessed the performance of \tool and the baselines on the problem of One-Step Dynamic Slicing.
In this task, we focus a fine-grained task on generating a single step of the slicing result.
Specifically, given a focal tracing step and a focal variable instance, we evaluate the effectiveness of identifying the last location where the variable instance was assigned.
Comparing with the full slicing set, the one-step slicing is more accurate and more useful in scenarios where developers need immediate, precise insights into the direct dependencies affecting a specific statement, allowing for efficient debugging and comprehension of complex program behavior without being overwhelmed by the entirety of the dynamic slice.
Additionally, if one-step slicing is executed with precision, then the aggregate slicing set will inherently be effective, as the accuracy of each individual step ensures that the complete dynamic slice comprehensively captures all pertinent influences on the point of interest.

\subsubsection{Benchmark}\label{sec:rq1benchmark}

We evaluated \tool\ and baselines on the following datasets:
\begin{itemize}[leftmargin=*]
    \item \textbf{ND-Slicer Translated Dataset.} Based on ND-Slicer~\cite{yadavally2024predictive}, which provides slicing results for Python programs, we translated the dataset to Java using an LLM, and carefully verified the semantic equivalence by comparing outputs between original and translated programs to ensure fairness. Since ND-Slicer outputs a sequence of slice steps, we adopted the first steps after the slicing criterion for our evaluation. This dataset contains \NDSlicerBenchmarkSize data points.
    \item \textbf{LLM-Slicer Dataset.} For LLM-Slicer~\cite{shahandashti2024program}, we used its \LLMSlicerBenchmarkSize benchmark programs for dynamic slicing. As LLM-Slicer predicts all slice steps (which may not match true execution order), we instrumented the programs to generate ground truths, and used heuristics when instrumentation was not possible.
    \item \textbf{\llmGeneratedDataset.} To better reflect real-world software complexity, we created the \llmGeneratedDataset, consisting of programs with library calls, complex control flows, and realistic class structures. The dataset is organized into five levels of increasing difficulty: \llmGeneratedSimple\ (basic API write-read pairs), \llmGeneratedComplex\ (adds unrelated code), \llmGeneratedAlias\ (introduces variable aliasing), \llmGeneratedMultiMethods\ (multiple methods), and \llmGeneratedMultiFiles\ (distributed across multiple files). Each level was systematically enhanced using LLMs to introduce greater complexity, and contains 1473 programs. 
    % Examples from each level are shown in \autoref{fig:datasetexample}.
    Due to space limit, detailed examples of each level is available in \cite{recov-slicing}.
\end{itemize}

\definecolor{codegreen}{rgb}{0,0.6,0}
\definecolor{codegray}{rgb}{0.5,0.5,0.5}
\definecolor{codepurple}{rgb}{0.58,0,0.82}
\definecolor{backcolour}{rgb}{0.95,0.95,0.92}

\lstdefinestyle{mystyle}{
    % backgroundcolor=\color{backcolour},   
    commentstyle=\color{codegreen},
    keywordstyle=\color{magenta},
    numberstyle=\tiny\color{codegray},
    stringstyle=\color{codepurple},
    basicstyle=\ttfamily\footnotesize,
    breakatwhitespace=false,         
    breaklines=true,                 
    captionpos=b,                    
    keepspaces=true,                 
    numbersep=5pt,                  
    showspaces=false,                
    showstringspaces=false,
    showtabs=false,                  
    tabsize=2,
    lineskip=0pt,
    columns=fullflexible,
    % numbers=none,
    % frame=none,
}

\lstset{style=mystyle}

\newsavebox{\llmcodesimple}
\begin{lrbox}{\llmcodesimple}
\begin{minipage}{0.45\textwidth}
\begin{lstlisting}[language=Java,escapechar=!]
public static void main(String[] args) {
    LinkedList<Integer> list = new LinkedList<>();
    /* write */ !\colorbox{yellow}{list.addFirst(42);}!
    /* read */ !\colorbox{green}{int value = list.getFirst();}!
    System.out.println(value);
}
\end{lstlisting}
\end{minipage}
\end{lrbox}

\newsavebox{\llmcodecomplex}
\begin{lrbox}{\llmcodecomplex}
\begin{minipage}{0.45\textwidth}
\begin{lstlisting}[language=Java,escapechar=!]
public static void main(String[] args) {
    LinkedList<Integer> list = new LinkedList<>();
    ArrayList<Integer> arrayList = new ArrayList<>();
    Stack<Double> stack = new Stack<>();
    stack.push(3.14);
    ...... // other unrelated code, 35 lines
    /* write */ !\colorbox{yellow}{list.addFirst(42);}!
    ...... // other unrelated code, 14 lines
    /* read */ !\colorbox{green}{int value = list.getFirst();}!
    System.out.println(value);
}
\end{lstlisting}
\end{minipage}
\end{lrbox}

% \begin{figure}
%     \subfloat[\textit{Simple}. Directed data dependency pair.]{\usebox{\llmcodesimple}}
    
%     \subfloat[\textit{Complex}. Adding unrelated lines.]{\usebox{\llmcodecomplex}}
% 	\caption{Code snippet from \llmGeneratedDataset of different difficulty levels. Each program is generated by add extra constructs to pervious difficulty level.}
% 	\label{fig:datasetexample}
% \end{figure}

\autoref{tab:rq1_dataset} provides statistics for the benchmark datasets used in this evaluation.
Comparing with the existing datasets, the \llmGeneratedDataset is more challenging, because it contains more complexity in number of lines and number of files.

\begin{table}
	\centering
	\caption{Benchmark Dataset Statistics}
	\label{tab:rq1_dataset}
    \resizebox{\columnwidth}{!}{
	\begin{tabular}{l|c|c|c}
		\toprule[1.2pt]
		\textbf{Dataset}     & \textbf{Size}           & \textbf{Avg. LOC} & \textbf{Avg. \#Files} \\
		\midrule
		ND-Slicer Translated & \NDSlicerBenchmarkSize  & 21.45               & 1                   \\
		LLM-Slicer           & \LLMSlicerBenchmarkSize & 49.35               & 1                   \\
        \llmGeneratedDataset Overall & 7365 & 90.17 & 1.40\\
        \midrule
		\llmGeneratedDataset \llmGeneratedSimple & 1473                     & 16.07               & 1                   \\
        \llmGeneratedDataset \llmGeneratedComplex & 1473                     & 97.47               & 1                   \\
        \llmGeneratedDataset \llmGeneratedAlias & 1473                     & 112.42               & 1                   \\
        \llmGeneratedDataset \llmGeneratedMultiMethods & 1473                     & 107.29               & 1                   \\
        \llmGeneratedDataset \llmGeneratedMultiFiles & 1473                     & 117.58               & 2.99                  \\
		\bottomrule[1.2pt]
	\end{tabular}
    }
\end{table}

\subsubsection{Baselines}

% We compared \tool with existing state-of-the-art slicing tools, including the trace-and-summary-based dynamic slicing tool, Slicer4J~\cite{ahmed2021slicer4j}, as well as neural-based dynamic slicing tools, ND-Slicer~\cite{yadavally2024predictive}, and LLM-Slicer~\cite{shahandashti2024program}. Additionally, we incorporated a re-execution-based approach. Slicer4J dynamically collects execution traces for application code and employs summary techniques for library code. ND-Slicer and LLM-Slicer use pre-trained neural networks or large language models to predict slices. Furthermore, the re-execution-based approach involves re-executing the program while instrumenting additional methods to capture missed dependencies.

% To ensure these baselines could function with our benchmark, we made the following modifications:
% \begin{itemize}[leftmargin=*]
% 	\item \textbf{Slicer4J}. We invoked Slicer4J with the command line option \texttt{-once} to generate one-step slicing results.
% 	\item \textbf{ND-Slicer}. Originally trained for Python, ND-Slicer required the training dataset to be translated from Python to Java using an LLM, while keeping the slicing results consistent. We then trained a Java version of ND-Slicer and evaluated it using our benchmark. Since ND-Slicer outputs all data dependencies, we utilize the first three elements to compute the percision and recall. To make a fair comparsion, if any of the top three is correct, we consider accuracy to be 100.
% 	\item \textbf{LLM Slicer}. We slightly modified the prompt to generate only one-step slicing results.
% \end{itemize}
We compared \tool with state-of-the-art dynamic slicing tools, including Slicer4J~\cite{ahmed2021slicer4j} (trace-and-summary-based), ND-Slicer~\cite{yadavally2024predictive} and LLM-Slicer~\cite{shahandashti2024program} (neural-based), as well as a re-execution-based baseline. Slicer4J collects dynamic execution traces for application code and summarizes library code. ND-Slicer and LLM-Slicer employ pre-trained neural models or LLMs to predict slices. The re-execution baseline re-runs the program with extra instrumentation to recover missed dependencies.

To adapt these tools for our benchmarks, we made the following adjustments:
\begin{itemize}[leftmargin=*]
    \item \textbf{Slicer4J}: Used with the \texttt{-once} flag for one-step slicing.
    \item \textbf{ND-Slicer}: As ND-Slicer was designed for Python, we translated its training data to Java via LLMs and trained a Java version. Since ND-Slicer outputs all dependencies, we evaluated the top three predictions; if any are correct, precision is 100\%.
    \item \textbf{LLM-Slicer}: Modified the prompt to output only one-step slices.
    \item \textbf{Re-execution Slicer}: Implemented using Microbat, this baseline re-executes programs with added instrumentation to capture additional data dependencies. Limitations include difficulties handling internal or native methods. We performed static analysis to locate possible dependency sites, then re-executed the code to capture missed dependencies.
\end{itemize}

\subsubsection{Experimental Results}

As shown in \autoref{tab:rq1}, RecovSlicing consistently outperforms all baselines across three dynamic slicing benchmarks. On \llmGeneratedDataset, RecovSlicing achieves a precision and recall of 80.34\%, substantially surpassing the best baseline (LLM Slicer: 39.01\% precision, 53.41\% recall). Notably, as detailed in \autoref{tab:rq1_dataset}, \llmGeneratedDataset is characterized by more complex program structures, including higher average lines of code and multiple files per program, making it significantly more challenging than the other datasets. The superior performance of RecovSlicing on this high-difficulty benchmark further highlights its robustness and effectiveness over existing approaches. 

On the ND-Slicer Dataset, RecovSlicing obtains 91.06\% for both precision and recall, outperforming ND-Slicer and LLM Slicer by more than 9 percentage points. For the LLM Slicer Dataset, RecovSlicing attains the highest scores with 98.25\% precision and recall, demonstrating robust and reliable dynamic slicing performance across diverse benchmarks. These results clearly demonstrate the effectiveness and superiority of RecovSlicing in both precision and recall, especially on more challenging datasets.

Compared to the re-execution-based approach, RecovSlicing achieves substantial improvements in both precision and recall on all benchmarks. For example, on the LLM Generated Dataset, RecovSlicing obtains 80.34\% precision and recall, whereas the re-execution method only reaches 5.03\%. This significant margin demonstrates that RecovSlicing is far more effective at accurately identifying slicing variables and their dependencies, especially in complex and large-scale programs. The poor performance of the re-execution baseline is primarily due to the fact that, in practice, many programs require instrumentation of a large number of classes during finding data dependencies. This often leads to timeouts or out-of-memory errors when performing re-execution, resulting in a high rate of failed cases and significantly lower overall effectiveness.

\begin{table}[t]
\centering\
\scriptsize
\caption{Quantitative Results on Dynamic Slicing Performance (RQ1).}
	\begin{tabular}{c|l|c|c}
		\toprule[1.2pt]
        \multirow{2}{*}{\textbf{Benchmark}} &\multicolumn{1}{c|}{\multirow{2}{*}{\textbf{Tool}}} & \multicolumn{2}{c}{\textbf{Evaluation Metrics}}\\
            \cline{3-4}
		& &\textbf{Precision(\%)} $\uparrow$ &\textbf{Recall(\%)} $\uparrow$ \\
        \hline
       \multirow{4}{*}{\textbf{LLM Generated Dataset}} 
        &Slicer4j & 11.09 & 32.08 \\
        &ND-Slicer & 12.90 & 33.20 \\ 
        &LLM Slicer & 39.01 & 53.41 \\ 
        &  Re-Execution Slicer &5.03  &5.03  \\ 
        & \cellcolor[rgb]{0.902,0.902,0.902}\textbf{RecovSlicing (ours)} &\cellcolor[rgb]{0.902,0.902,0.902}\textbf{80.34} &\cellcolor[rgb]{0.902,0.902,0.902}\textbf{80.34} \\  
         \hline
        \multirow{4}{*}{\textbf{\makecell{ND-Slicer Dataset \\(translated to Java)}}}
         &Slicer4j & 71.30 & 56.62 \\
         &  ND-Slicer & 82.00 & 79.10 \\ 
         &  LLM Slicer & 81.07 & 76.05 \\ 
         & Re-Execution Slicer &79.1  &47.39  \\
         & \cellcolor[rgb]{0.902,0.902,0.902}\textbf{RecovSlicing (ours)} &\cellcolor[rgb]{0.902,0.902,0.902}\textbf{91.06} &\cellcolor[rgb]{0.902,0.902,0.902}\textbf{91.06} \\  
        \hline
        \multirow{4}{*}{\textbf{LLM Slicer Dataset}}
         & Slicer4j & 20.54 & 80.77 \\
         &  ND-Slicer & 59.86 & 87.13 \\  
         &  LLM Slicer & 32.63 & 38.01 \\ 
         &  Re-Execution Slicer &26.17  &10.09  \\
         & \cellcolor[rgb]{0.902,0.902,0.902}\textbf{RecovSlicing (ours)} &\cellcolor[rgb]{0.902,0.902,0.902}\textbf{98.25} &\cellcolor[rgb]{0.902,0.902,0.902}\textbf{98.25} \\
		\bottomrule[1.2pt]
	\end{tabular}
	\label{tab:rq1}
\end{table}

% \begin{figure}[tbp]
%   \centering
%   \subfloat[Percision]
%   {\includegraphics[width=1.0\linewidth]
%   {image/RQ1-precision-for-each-dataset.pdf}}
%   \vspace{-0.4cm}
%   \subfloat[Recall]{\includegraphics[width=1.0\linewidth]{image/RQ1-recall-for-each-dataset.pdf}}  
%   \caption{Comparing Different Difficulty Levels.}
%   \label{fig:RQ1-recall-for-each-dataset}
% \end{figure}

\begin{figure}[tbp]
  \centering
{\includegraphics[width=0.9\linewidth]{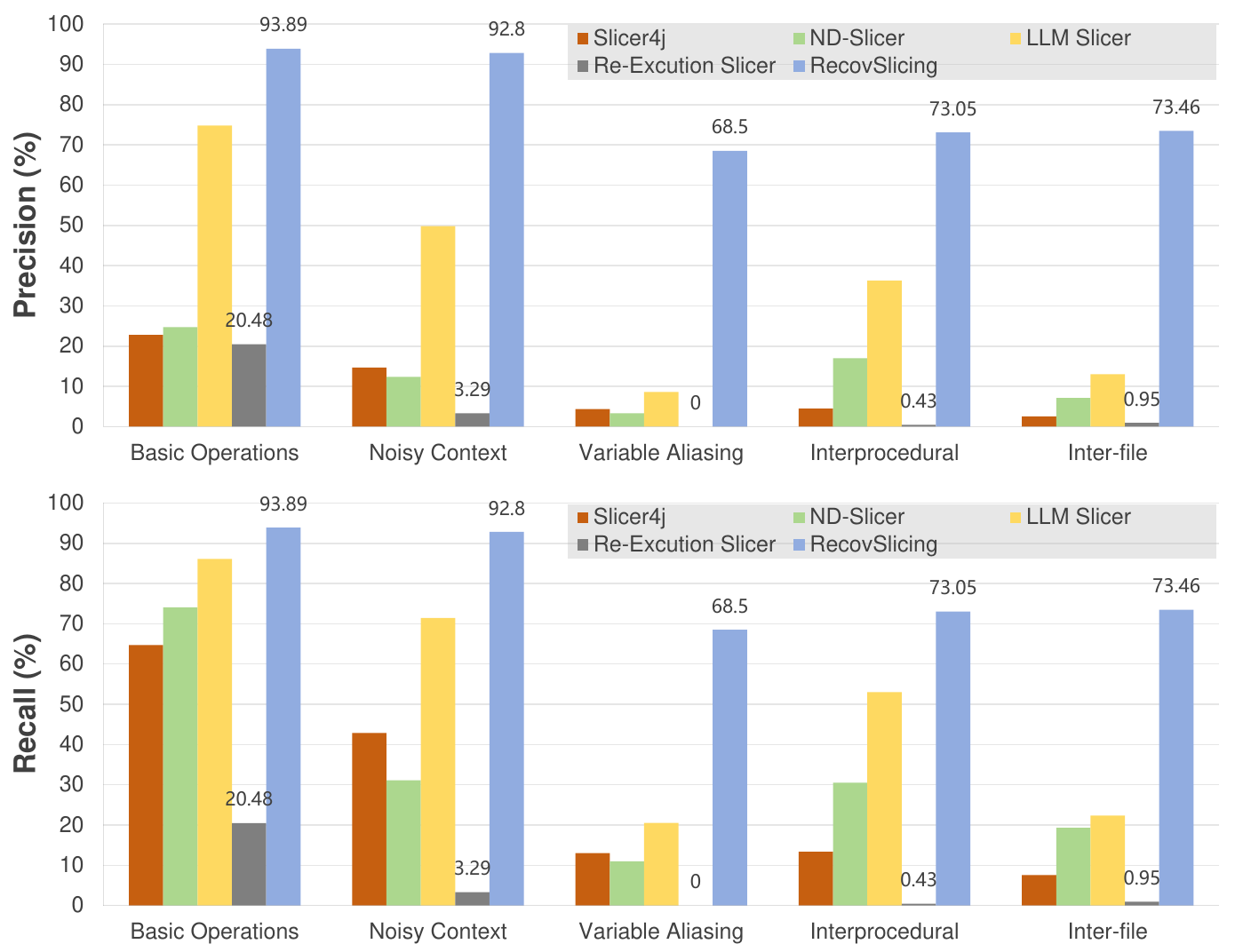}}  
  \caption{Comparing Different Difficulty Levels.}
  \label{fig:RQ1-recall-for-each-dataset}
\end{figure}

% \begin{table}[h]
% \centering
% \caption{RQ1 results \hongshu{add caption}}
% \label{tab:rq1}
% \input{tables/rq1_table}
% \end{table}

% \begin{table}
% 	\centering
% 	\caption{Comparison of Re-Execution Slicer and \tool on \llmGeneratedDataset}
% 	\label{tab:rq1_reexecution}
% 	\begin{tabular}{l|c|c}
% 		\hline
% 		\textbf{Tool}       & \textbf{Accuracy} & \textbf{Recall} \\
% 		\hline
% 		Re-Execution Slicer & 26.17             & 10.01           \\
% 		\tool               & \textbf{67.25}    & \textbf{85.26}  \\
% 		\hline
% 	\end{tabular}
% \end{table}

% \begin{table}[t]
% \centering\
% % \scriptsize
% \caption{Comparison of Re-Execution Slicer and \tool on \llmGeneratedDataset}
% 	\begin{tabular}{l|c|c}
% 		\toprule[1.2pt]
% 		\textbf{Tool}       & \textbf{Accuracy} & \textbf{Recall} \\
% 		\hline
% 		Re-Execution Slicer & 26.17             & 10.01           \\
% 		\tool               & \textbf{67.25}    & \textbf{85.26}  \\
% 		\bottomrule[1.2pt]
% 	\end{tabular}
% 	\label{tab:rq1_reexecution}
% \end{table}

\textbf{Comparing Different Difficulty Levels.} \autoref{fig:RQ1-recall-for-each-dataset} show the result of different difficulty levels on \llmGeneratedDataset. In general, as task difficulty increases, both the precision and recall of all models decrease. \tool consistently maintains an advantage over the baseline, particularly on tasks of higher difficulty. Furthermore, we observe that the introduction of aliasing causes a performance decline in all models, suggesting that aliasing poses additional challenges for data dependency analysis.

\textbf{Comparing with Local Models.} To enhance the applicability and deployment flexibility of our approach and ensure data security, we evaluated \tool using both GPT-4o and a locally deployable open-source model with 27b parameters, \textit{unsloth/gemma-3-27b-it-unsloth-bnb-4bit}. In \autoref{tab:gemma}, experimental results indicate that, while the use of the local model results in a moderate reduction in accuracy, its performance remains superior to that of the baseline.

\begin{table}[tbp]
    \centering
    \caption{Comparing between Remote Models and Local Models}
    \label{tab:gemma}
    \resizebox{\columnwidth}{!}{
    \begin{tabular}{l|c|c|c|c|c}
         \toprule[1.2pt]
         & \llmGeneratedSimple & \llmGeneratedComplex & \llmGeneratedAlias & \llmGeneratedMultiMethods & \llmGeneratedMultiFiles \\
         \midrule
         \textbf{GPT-4o} & 93.89 & 92.8 & 68.50 & 73.05 & 73.46 \\
         \textbf{Gemma 3 27b} & 77.96 & 72.72 & 55.22 & 69.26 & 70.82 \\
         \bottomrule[1.2pt]
    \end{tabular}}
\end{table}

% \subsubsection{Case Analysis}
% We analyzed cases from the \llmGeneratedDataset to understand \tool's performance. Figure \autoref{fig:case1} shows success case.
% In this snippet, the program performs various operations on a list object. In the first line, the program add the first element to the list and in the third line, the program read the first element of the list. Between the two lines, there are multiple lines omitted, which performs extra adding, removing operations but not to the first element.  We requested \tool and the baselines to generate one-step slicing results for this code snippet.
% \tool successfully ignored other \texttt{add} operations and returned the correct slicing result, by clearly utilizing the slicing criteria of \texttt{list.elementData[0]}. In contrast, the baselines failed to generate the correct slicing result. Slicer4J, ND-Slicer, and LLM-Slicer incorrectly returned middle lines. The re-execution-based approach failed to generate any slicing result.

\subsubsection{Case Analysis}
We further analyzed cases from the \llmGeneratedDataset to better understand \tool's performance. As illustrated in \ref{fig:case1}, the program performs several operations on a list object. The first line adds an element to the list, while the third line reads the first element; intermediate lines perform unrelated modifications. We asked \tool and all baselines to generate one-step slices for this snippet. \tool correctly identified the relevant slicing criterion (\texttt{list.elementData[0]}), ignoring irrelevant operations, and produced the precise slice. In contrast, Slicer4J, ND-Slicer, and LLM-Slicer mistakenly included unrelated lines, while the re-execution-based approach failed to generate any result.

\definecolor{codegreen}{rgb}{0,0.6,0}
\definecolor{codegray}{rgb}{0.5,0.5,0.5}
\definecolor{codepurple}{rgb}{0.58,0,0.82}
\definecolor{backcolour}{rgb}{0.95,0.95,0.92}

\newsavebox{\llmcodecasefirst}
\begin{lrbox}{\llmcodecasefirst}
\begin{minipage}{0.45\textwidth}
\begin{lstlisting}[language=Java,escapechar=!]
/* write */ list.addFirst(1);
// other operations omitted (e.g. list.add(2);) ...
/* read */ int value = list.getFirst();
\end{lstlisting}
\end{minipage}
\end{lrbox}

\newsavebox{\llmcodecasesecond}
\begin{lrbox}{\llmcodecasesecond}
\begin{minipage}{0.45\textwidth}
\begin{lstlisting}[language=Java,escapechar=!]
WeakHashMap<String, String> map = new WeakHashMap<>();
String key = new String("key");
String value = "value";
/* write */ map.put(key, value);
/* read */ String readValue = map.get(key);
\end{lstlisting}
\end{minipage}
\end{lrbox}

\lstdefinestyle{mystyle}{
    backgroundcolor=\color{backcolour},   
    commentstyle=\color{codegreen},
    keywordstyle=\color{magenta},
    numberstyle=\tiny\color{codegray},
    stringstyle=\color{codepurple},
    basicstyle=\ttfamily\footnotesize,
    breakatwhitespace=false,         
    breaklines=true,                 
    captionpos=b,                    
    keepspaces=true,                 
    numbers=left,                    
    numbersep=5pt,                  
    showspaces=false,                
    showstringspaces=false,
    showtabs=false,                  
    tabsize=2
}

\lstset{style=mystyle}

\begin{figure}
	\subfloat[Success Case\label{fig:case1}]{\usebox{\llmcodecasefirst}}
    
    \subfloat[Failure Case\label{fig:case2}]{\usebox{\llmcodecasesecond}}
	\caption{Code snippets from \llmGeneratedDataset. Due to space limit, we only show partial programs. The full program is available in \cite{recov-slicing}.}
\end{figure}

We also examined failure cases. While LLMs are generally reliable, there are scenarios where they make incorrect predictions. For example, as shown in Figure \ref{fig:case2}, the query slicing criterion is \texttt{map.table[14].hash}. Here, the LLM fails to accurately identify the variable definition, as it cannot determine which map entry is stored at index 14 due to insufficient contextual information.

Due to space limit, more detailed examples and results are available in our repository\cite{recov-slicing}.

\vspace{0.5ex}
\finding{\textbf{RQ1: }\tool outperforms existing slicing tools in the one-step dynamic slicing task, demonstrating its effectiveness in accurately predicting slices and recovering true positives. Specifically, \tool is more accurate on tracking one slicing variable and its dependencies, making it particularly useful for debugging and comprehension tasks.}

\subsection{Component-Wise Analysis (RQ2)}

% In RQ2, we evaluate the performance of \tool in its two main components: variable recovery and definition inference. We utilize exhaustive instrumentation to obtain the ground truth data for this evaluation, and then compare the component results of \tool against this ground truth.

In RQ2, we evaluate the performance of \tool's two core components: variable recovery and definition inference. Ground truth is obtained via exhaustive instrumentation, enabling direct comparison with \tool's outputs.

\subsubsection{Setup}
% Since exhaustive instrumentation is required to obtain the ground truth, we selected a benchmark consisting of 450 programs from the \llmGeneratedDataset, which are of medium size and can be fully instrumented. Additionally, we conducted a detailed analysis of 28 programs from the official LeetCode \cite{leetcode} solutions. For each program, we generated a series of variable expansion tasks and definition inference tasks to enhance the benchmark's diversity. Overall, we constructed a dataset comprising 570 variable expansion tasks (covering 47 data types from the Java standard library) and 1032 definition inference tasks. The dataset statistics are presented in \autoref{tab:rq2_dataset}.
% We select 450 fully instrumentable, medium-sized programs from the \llmGeneratedDataset, and 28 representative programs from official LeetCode solutions~\cite{leetcode}. For each program, we systematically generate variable expansion and definition inference tasks to ensure diversity. In total, the benchmark consists of 570 variable expansion tasks—spanning 47 standard Java data types—and 1032 definition inference tasks. Detailed dataset statistics are summarized in \autoref{tab:rq2_dataset}.
To enable full instrumentation, which is only feasible for programs of moderate size, we select 450 medium-sized programs from the \llmGeneratedDataset. Additionally, we include 28 representative programs from official LeetCode solutions~\cite{leetcode}. For each program, we systematically construct both variable recovery and definition inference tasks to maximize benchmark diversity. In total, the dataset comprises 570 variable recovery tasks (covering 47 standard Java data types) and 1032 definition inference tasks. 
% Dataset statistics are summarized in \autoref{tab:rq2_dataset}.

% \begin{table}[t]
% \centering\
% % \footnotesize
% \caption{Component-Wise Analysis Dataset}
% \resizebox{\columnwidth}{!}{
% 	\begin{tabular}{l|c|c|c}
% 		\toprule
%         \textbf{Dataset}                              & LeetCode & \llmGeneratedDataset & Total \\
% 		% \hline
%         \midrule
% 		\textbf{Num. of Programs}                   & 28            & 450                       & 478   \\
% 		\textbf{Num. of Variable Expansion Tasks}   & 120            & 450                        & 570   \\
% 		\textbf{Num. of Definition Inference Tasks} & 350            & 682                        & 1032  \\
% 		\bottomrule
% 	\end{tabular}
%     }
% 	\label{tab:rq2_dataset}
% \end{table}

\subsubsection{Measurement}
We evaluate variable recovery and definition inference using the following metrics:

\textit{Variable Recovery.}
We use precision and recall to assess variable recovery. For each variable instance $var$ with $k$ ground-truth children, if \tool recovers $c$ children, and $x$ of them match the ground-truth in both variable name and value, then
$\mathrm{Precision} = x/c, \mathrm{Recall} = x/k$.

\textit{Definition Inference.}
Definition inference is measured by the success ratio. For each read step $s$ with ground-truth defining step $s_{def}$, \tool is successful if it correctly identifies $s_{def}$. Let $d$ be the total number of data dependencies, and $x$ the number correctly recovered, then
$\mathrm{Success\ Ratio} = x/d$.

% \begin{table}[t]
% 	\centering
% 	% reduce column spacing
% 	\setlength{\tabcolsep}{5.5pt}
% 	\caption{Quantitative Results on Value Prediction (RQ2).}
% 	\begin{tabular}{c | c |c |c }
% 		\hline
% 		% \multicolumn{2}{c|}{\textbf{Benchmark}} & \textbf{Tool} & \textbf{Accuracy (\%)} $\uparrow$ & \textbf{Recall (\%)} $\uparrow$ \\
% 		\multirow{2}{*}{\textbf{Dataset}}  & \multicolumn{2}{c|}{\textbf{Variable Expansion}} & \textbf{Definition Inference}                                \\
% 		\cline{2-4}
% 		                                   & \textbf{Precision (var)}                         & \textbf{Recall (var)}         & \textbf{Success Ratio (def)} \\
% 		\hline
% 		\textbf{From LeetCode}             & 86.14\%                                          & 77.56\%                       & 74.86\%                      \\
% 		\textbf{From \llmGeneratedDataset} & 92.57\%                                          & 91.92\%                       & 87.54\%                      \\
% 		\textbf{Total}                     & 91.48\%                                          & 89.24\%                       & 83.24\%                      \\
% 		\hline
% 	\end{tabular}
% 	\label{tab:rq2}
% \end{table}

\begin{table}[t]
\centering
% \footnotesize
\caption{Quantitative Results on Value Prediction (RQ2).}
\label{tab:rq2}
\resizebox{\columnwidth}{!}{
	\begin{tabular}{lccc}
		\toprule[1.2pt]
		\multirow{2}{*}{\textbf{Dataset}}  & \multicolumn{2}{c }{\textbf{Variable Recovery}} & \textbf{Definition Inference}                                \\
		\cmidrule(l){2-3} \cmidrule(l){4-4}
		                                   & \textbf{Precision}                         & \textbf{Recall}         & \textbf{Success Ratio} \\
		\midrule
		\textbf{LeetCode}             & 86.14\%                                          & 77.56\%                       & 74.86\%                      \\
		\textbf{\llmGeneratedDataset} & 92.57\%                                          & 91.92\%                       & 87.54\%                      \\
		\textbf{Total}                     & 91.48\%                                          & 89.24\%                       & 83.24\%                      \\
		\bottomrule[1.2pt]
	\end{tabular}
    }
	
\end{table}

\subsubsection{Experimental Results}
As shown in \autoref{tab:rq2}, \tool achieves a precision of 91.48\% and a recall of 89.24\% in variable recovery tasks. Our investigation indicates that some false positives and false negatives are largely due to an incomplete data store, which leads to failures in variable recovery. More qualitative examples can be found in \cite{recov-slicing}.

% \tool achieves the success ratio of 83.24\% in the task of definition inference. Although LLMs are generally effective at recovering execution information, they may produce hallucinations when crucial context information is missing from the prompt. \autoref{fig:neg-example} presents an example where \tool fails to successfully infer a definition. In this case, when a variable of type \texttt{HashSet} performs an \textit{add} operation, the LLM is asked to determine whether a local field, \texttt{set.map.table.table[4]}, is written to or modified. While this inference is indeed correct, arriving at this conclusion requires an understanding of the internal mapping rules of \texttt{HashSet}. Without more detailed context, the LLM struggles to make an accurate prediction, ultimately producing an incorrect answer. In the future work, we will explore the technique to search for and query more relevant
% contextual information to further improve the recovering performance.

% \begin{figure}
% 	\centering
% 	\includegraphics[width=1.0\linewidth]{image/neg-example.pdf}
% 	\caption{Example where \tool fails to infer a data definition.\label{fig:neg-example}}
% \end{figure}

\finding{\textbf{RQ2:} \tool achieves a precision of 91.48\% and a recall of 89.24\% in variable recovery tasks, and a success ratio of 83.24\% in definition inference tasks.}
\subsection{Ablation Study (RQ3)}

In this experiment, we quantify the impact of the in-context learning component within \tool on overall performance. To this end, we compare \tool’s full version (denoted as w/ IC) with a variant that disables in-context learning (denoted as w/o IC). Both variants are evaluated on the benchmarks described in Section~\ref{sec:rq1benchmark}, with results summarized in \autoref{tab:rq3}.

As shown in the table, incorporating in-context learning consistently improves accuracy across all benchmarks. Even in simpler or less noisy cases, in-context learning leads to improvements of 5-7\%. Overall, these results demonstrate that in-context learning is particularly beneficial for handling complex program features and noisy contexts.

The design of in-context learning in \tool specifically targets the improvement of key steps, such as variable recovery and definition inference. As discussed in RQ2, these components already demonstrate high precision and recall. The results in \autoref{tab:rq3} further highlight that in-context learning is especially effective for complex scenarios, such as those involving library calls, intricate control flows, and multi-file programs in the LLM Generated Dataset, leading to significant end-to-end performance gains.

\begin{table}[t]
\centering\
\footnotesize
\caption{Quantitative Results on Ablation Study  (RQ3).}\label{tab:rq3}

\begin{tabular}{lcc}
    \toprule[1.2pt]
    \multirow{2}{*}{\textbf{Benchmark}} & \multicolumn{2}{c}{\textbf{Accuracy (\%)}} \\
    \cmidrule{2-3}
    & \textbf{w/ IC} & \textbf{w/o IC} \\
    \midrule
    Basic Operations & \textbf{93.89} & 87.64 \\
    Noisy Context & \textbf{92.80} & 85.61 \\
    Variable Aliasing & \textbf{68.50} & 43.46 \\
    Interprocedural & \textbf{73.05} & 68.97 \\
    Inter-file & \textbf{73.46} & 71.08 \\
    \midrule
    ND-Slicer Dataset & \textbf{91.06} & 79.49 \\
    LLM Slicer Dataset & \textbf{98.25} & 93.18 \\
    \bottomrule[1.2pt]
\end{tabular}
\end{table}

\finding{\textbf{RQ3:} The in-context learning component significantly improves the overall performance of \tool.}
\subsection{Application: Debugging Performance (RQ4)}

\begin{figure}[t]
  \centering
  \includegraphics[width=0.65\linewidth]{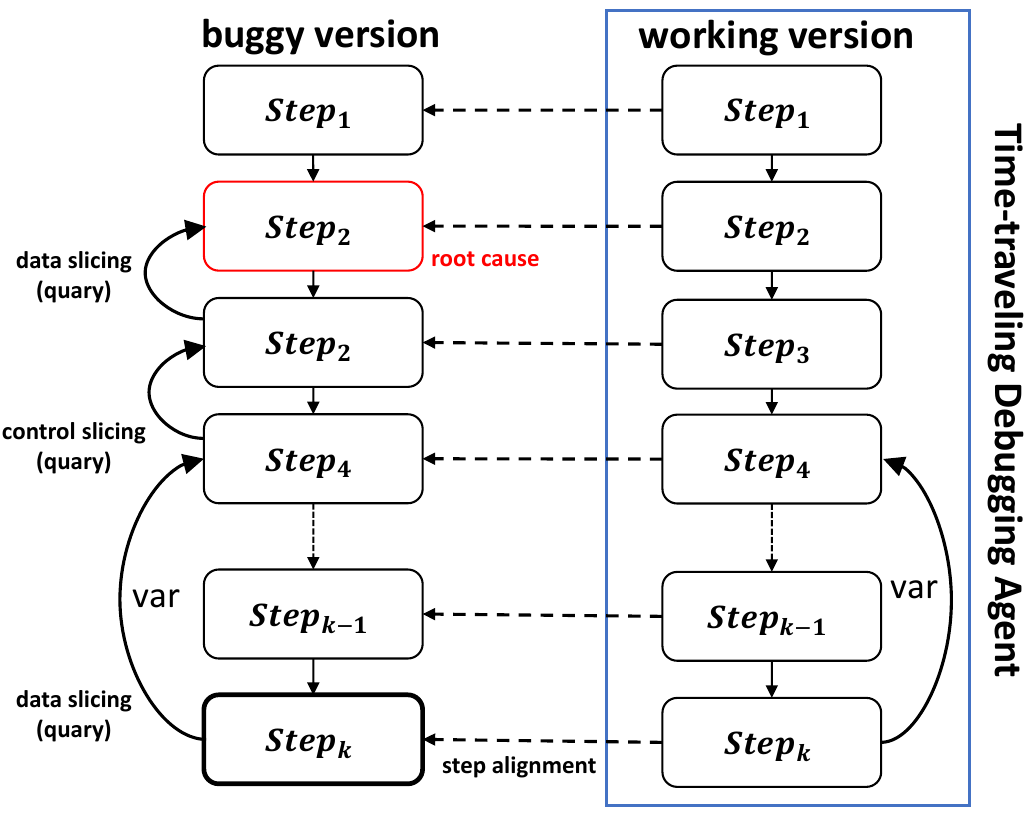}
  \caption{
  The debugging agent used in this experiment.
  % Taking the trace of the working version (i.e., $tr_w$) as the reference,
  % the time-travelling debugging agent can issue queries (as control and data slicing) on the steps of the trace of buggy version (i.e., $tr_b$),
  % gradually moving toward the root-cause step on the buggy trace.
  % Specifically, if $s\in tr_b$ can be aligned with $s'\in tr_w$ but $s$ and $s'$ read variable $var$ of different runtime value, the debugging agent can issue a data slicing on $s$ regarding $var$;
  % if $s\in tr_b$ cannot be aligned with any steps in $tr_w$, the debugging agent can issue a control slicing on $s$;
  % if $s\in tr_b$ can be aligned with $s'\in tr_w$ and all their read variables share the same runtime value, there is no query can be generated.
  }\label{fig:aligned-trace-example1}
\end{figure}

\begin{sloppypar}
Time-travelling debugging represents a specialized debugging methodology in which the complete execution trace of a program is recorded. This enables debugging to proceed by navigating the trace backwards from the point where an anomaly is observed (such as an exception being thrown or an incorrect value being detected), step by step, to ultimately identify the root cause of the bug. For instance, as illustrated in \autoref{fig:aligned-trace-example1}, two traces are shown: in the left trace, an anomaly is observed at $step_k$, which is actually caused by an earlier code error at $step_2$, but $step_2$ is unknown. We refer to the code error location as root cause. The objective of a debugging agent is to start from $step_k$ and identify a path through the trace that leads to the root cause, where the causal error occurred. Each edge in this path corresponds to a data or control dependency, and if the debugging agent can successfully identify the root cause $step_2$, the debugging is considered successful.
\end{sloppypar}

In particular, our experiment focuses on a regression debugging scenario, where two different but similar program versions are available: one trace exhibits successful execution (the working version), while the other trace corresponds to a failing execution (the buggy version). The current state-of-the-art debugging agent in this context is \debuggingtool \cite{wang2019explaining}, which operates by aligning the traces of both versions and identifying inconsistencies to localize the root cause of the regression bug.

Due to the complexity of real-world projects, \debuggingtool often misses data dependencies because certain traces are not recorded. In this evaluation, we integrate our tool RecovSlicing, into \debuggingtool: when searching for data dependencies, we additionally employ RecovSlicing to leverage the LLM to recover potential missing data dependencies.

\subsubsection{Benchmark}

We use the Defects4J dataset as our benchmark. Out of the 835 bugs in Defects4J, we select 489 bugs for evaluation, as the state-of-the-art debugging agent \debuggingtool is limited in its ability to collect long traces.

% Given a passing test case in a project,
% we record its coverage and apply random mutations on the covered lines of code
% until the test case fails.
% We choose 8 types of mutations in this study, i.e.,
% removing if block,
% removing assignment,
% removing if condition,
% negating if condition,
% changing unary operator,
% changing binary operator,
% swapping the variable order, and
% changing the literals (including string and number literal).
% We also remove the cases where the trace is too long for \debuggingtool to process and
% the root cause step just lies in the end of trace (i.e., trivial case).
% As a result,
% we collect 1152 bugs
% from 12 Java projects.

\subsubsection{Implementation}

In this experiment, we instrument the application code, imposing a trace length limit of up to 100,000 steps. To account for the additional overhead from instrumentation and potential GPT network latency, we terminate any debugging session that exceeds 60 minutes in total runtime; such sessions are counted as failed debugging attempts. Instrumentation is configured to record variables up to one layer deep. We integrate \tool into the debugging agent, invoking it whenever the agent requires, but fails to access, certain data dependencies.

% Additionally, we curate a dataset comprising 169 examples for variable expansion and 658 examples for definition analysis. These examples are collected from widely used Java libraries (e.g., \texttt{ArrayList} and \texttt{HashMap}) and are manually checked to ensure there is no overlap with the subject programs.

% \subsubsection{Implementation}
% In this experiment, we let the instruments the application code,
% with a trace length limit up to 100,000 steps.
% In addition, considering that the additional instrumentation and GPT network latency can incur unexpected runtime cost,
% we will kill a debugging session if it takes over 60 minutes.
% %\textcolor{red}{Comment 9: trace generation is limited to 10 minutes. Overall debugging time is limited to 60 minutes (trace generation + LLM).}
% The killed session is counted as a failed debugging process.
% Further, we allow the instrumentation to record the variables with 1 layer.
% In addition, we integrate \tool into the debugging agent
% each time it is required but fail to access data dependencies.
% Last, we prepare a dataset of 169 examples of variable expansion
% and 658 examples of definition analysis.
% We collect the examples from the most popular used Java libraries (e.g., \texttt{ArrayList} and \texttt{HashMap}),
% which are manually checked to avoid overlapping with the subject programs.

\subsubsection{Measurement}

We measure the effectiveness of \tool to improve the performance of the time-travelling debugging by its debugging success rate.
Specifically, given a bug $b = \langle test, v_b, v_w \rangle$,
where $v_b$ represents the buggy version,
$v_w$ represents the working version, and
$test$ represents the test case which can pass on $v_w$ and fail on $v_b$.
Following the practice in \cite{wang2019explaining},
we call the debugging agent successful in $b$ if it can
generate the debugging process from the observable faulty step
to the root-cause step.

% \subsubsection{Baselines}
% To have more comprehensive executions,
% we can alternatively instrument more of the code under the debugging task.
% In the experiment,
% we choose different instrumentation strategies
% as the baselines.
% \begin{itemize}[leftmargin=*]
%   \item \textbf{Extra Library Instrumentation (ELI)}:
%     For each bug, we collect all its used Java libraries so that
%     many implicit internal program behaviors are now observable.
%   \item \textbf{Extra Variable Instrumentation (EVI)}:
%     For each variable of complex type,
%     we expand its children for $k$ layers,
%     so that more variables can be observable on the trace.
%     We set $k$ to be $1, 3$ and $5$,
%     denoted as EVI-$1$, EVI-$3$, and EVI-$5$.
% \end{itemize}

% Note that, in practice, there is always a tradeoff between
% observability and runtime cost and trace-exploring efforts.
% The experiment also allows us to investigate
% the cost-effectiveness of different strategies.

\subsubsection{Results}
\autoref{tab:rq4} shows the performance of integrating \tool to \debuggingtool on the 16 active projects in Defects4J, as well as the overall performance,
in contrast to \debuggingtool alone.
\tool outperforms or has the same performance as \debuggingtool on every project in Defects4J.
Overall, \tool allows the debugging agent to have the debugging success rate of 0.89 in the Defects4J dataset,
while \debuggingtool achieves 0.73.
Notably, \tool achieves a debugging success rate above 0.75 on 14 out of 16 projects, while \debuggingtool only achieves this success rate on 9 projects.
Therefore, we can conclude that by recovering miss-recorded data dependencies through \tool, the performance of downstream tasks such as time-travelling debugging will be improved.

\begin{table}[t]
\centering
\footnotesize
\caption{Quantitative Results on Debugging Performance (RQ4). 
    \textbf{\tool (ours)} outperforms \textbf{\debuggingtool (TR)} or achieves the same performance on every active project in Defects4J.}
	% \begin{tabular}{lccc}
	% 	\toprule[1.2pt]
 %        \multirow{2}{*}{\textbf{Project}} &\multirow{2}{*}{\textbf{Trace Length}} & \multicolumn{2}{c}{\textbf{Debugging Success Rate}}\\
 %            \cmidrule{3-4}
	% 	& &\textbf{Tregression} &\textbf{Ours}\\
 %        % \hline
 %        \midrule
 %        {Chart} & 5,434.50 & 0.85 & \textbf{0.90} \\
 %        % \hline
 %        {Cli} & 667.00 & 0.55 & \textbf{0.75} \\
 %        % \hline
 %        {Closure} & 59,330.00 & 0.65 & \textbf{0.90} \\
 %        % \hline
 %        {Codec} & 1,263.73 & 0.85 & \textbf{1.00} \\
 %        % \hline
 %        {Compress} & 2,611.07 & 0.71 & \textbf{0.84} \\
 %        % \hline
 %        {Csv} & 227.83 & 0.27 & \textbf{0.91} \\
 %        % \hline
 %        {Gson} & 523.00 & \textbf{0.92} & \textbf{0.92} \\
 %        % \hline
 %        {JacksonCore} & 2,021.14 & 0.85 & \textbf{1.00} \\
 %        % \hline
 %        {JacksonDatabind} & 5,319.16 & 0.60 & \textbf{0.66} \\
 %        % \hline
 %        {JacksonXml} & 792.33 & 0.50 & \textbf{0.67} \\
 %        % \hline
 %        {Jsoup} & 3,797.91 & 0.69 & \textbf{0.93} \\
 %        % \hline
 %        {JxPath} & 3,635.25 & 0.83 & \textbf{0.90} \\
 %        % \hline
 %        {Lang} & 1,438.18 & 0.80 & \textbf{0.94} \\
 %        % \hline
 %        {Math} & 4,862.13 & 0.90 & \textbf{0.96} \\
 %        % \hline
 %        {Mockito} & 2,658.50 & 0.83 & \textbf{1.00} \\
 %        % \hline
 %        {Time} & 1,649.75 & 0.81 & \textbf{0.90} \\
 %        % \hline
 %        \midrule
 %        \textbf{Overall} & 7,776.99 & 0.73 & \textbf{0.89} \\      
	% 	\bottomrule[1.2pt]
	% \end{tabular}
    \begin{tabular}{lccc}
    \toprule[1.2pt]
    \multirow{2}{*}{\textbf{Project}} & \multirow{2}{*}{\textbf{Trace Length}} & \multicolumn{2}{c}{\textbf{Debugging Success Rate}}\\
    \cmidrule{3-4}
    & & \textbf{Tregression} & \textbf{Ours}\\
    \midrule
    {Chart} & 5.43k & 0.85 & \textbf{0.90} \\
    {Cli} & 0.67k & 0.55 & \textbf{0.75} \\
    {Closure} & 59.33k & 0.65 & \textbf{0.90} \\
    {Codec} & 1.26k & 0.85 & \textbf{1.00} \\
    {Compress} & 2.61k & 0.71 & \textbf{0.84} \\
    {Csv} & 0.23k & 0.27 & \textbf{0.91} \\
    {Gson} & 0.52k & \textbf{0.92} & \textbf{0.92} \\
    {JacksonCore} & 2.02k & 0.85 & \textbf{1.00} \\
    {JacksonDatabind} & 5.32k & 0.60 & \textbf{0.66} \\
    {JacksonXml} & 0.79k & 0.50 & \textbf{0.67} \\
    {Jsoup} & 3.80k & 0.69 & \textbf{0.93} \\
    {JxPath} & 3.64k & 0.83 & \textbf{0.90} \\
    {Lang} & 1.44k & 0.80 & \textbf{0.94} \\
    {Math} & 4.86k & 0.90 & \textbf{0.96} \\
    {Mockito} & 2.66k & 0.83 & \textbf{1.00} \\
    {Time} & 1.65k & 0.81 & \textbf{0.90} \\
    \midrule
    \textbf{Overall} & 7.78k & 0.73 & \textbf{0.89} \\      
    \bottomrule[1.2pt]
\end{tabular}
	\label{tab:rq4}
\end{table}

% \begin{table}[h]

% \centering
% \caption{RQ4 results \hongshu{add caption}}
% \label{tab:rq4}

% \input{tables/rq4_table}

% \end{table}

\finding{\textbf{RQ4:} \tool improves debugging success rate from 0.73 to 0.89 in the real-world Defects4J dataset, demonstrating the effectiveness of \tool in enhancing the performance of time-travelling debugging agents on real-world programs.}

\section{Related Work}

\textbf{Dynamic Slicing.} Dynamic slicing \cite{korel1988dynamic, agrawal1990dynamic} analyzes execution-specific dependencies and is widely used in debugging and testing \cite{weiser1984program, agrawal1993debugging, barr2014tardis, lin2017feedback, harman1995using, binkley1998application}. Various enhancements \cite{venkatesh1991semantic, canfora1998conditioned, gupta1995hybrid, hall1995automatic, xu2005brief, silva2012vocabulary, li2020more} have been proposed over decades. A central challenge is the runtime overhead of collecting execution traces and constructing Program Dependence Graphs (PDG) \cite{zhang2004cost, zhang2005cost, zhang2006dynamic, wang2007hierarchical}.
To reduce overhead, researchers explored partial instrumentation, such as Call-Mark Slicing \cite{nishimatsu1999call} and checkpoint-assisted methods \cite{zhang2006dynamic}, which selectively log execution events. Others use precomputed dynamic dependency summaries \cite{palepu2013improving, palepu2017dynamic, ahmed2021slicer4j}, such as in Slicer4J \cite{ahmed2021slicer4j}, which summarizes library calls to avoid tracing their executions. However, such summaries are input-insensitive and cannot generalize to unseen libraries or runtime conditions. \tool instead leverages partial trace data and LLMs to approximate dependencies dynamically, addressing both accuracy and generalizability.

\textbf{Slicing without Execution.} Recent works apply neural models to predict slices from static code alone \cite{yadavally2023partial, yadavally2024predictive, shahandashti2024program, ding2024traced, wang2024llmsa}. ND-Slicer \cite{yadavally2024predictive} and LLM-Slicer \cite{shahandashti2024program} bypass execution tracing entirely by using transformer-based or general-purpose LLMs to infer dependency paths. However, these methods are limited by token length, lack of variable value inference, and poor performance on complex control flows or aliasing scenarios. \tool complements such models by incorporating runtime traces and performing adaptive step-wise recovery, enabling effective slicing even in non-deterministic or multi-library programs.

\textbf{LLMs for Program Analysis.} LLMs are increasingly applied in program analysis tasks such as bug detection \cite{fan2023large, li2024enhancing, mohajer2024effectiveness}, dependency reasoning \cite{ma2023lms, shahandashti2024program}, and static slicing \cite{wang2025contemporary, krishna2024codellm, fang2024large}. While promising, existing methods often lack dynamic context, and struggle to maintain accurate control/data flow across larger codebases. \tool differs by tightly integrating execution data with LLMs, improving both precision and robustness over purely static or prediction-only approaches.
% \section{Conclusion}

% In this work, we propose \tool, a novel approach effectively computes dynamic data dependency in programs using LLM-based inference and program trace analysis. 
% It achieves high precision and recall, outperforming state-of-the-art slicers such as Slicer4J and ND-Slicer.
% \tool mitigates alias inference challenges and eliminates the need for re-execution. 
% The setting is typically useful for practical debugging scenarios.
% In our future work, we will build a more sophisticated dynamic data dependency dataset and extend \tool in more programming languages. 

\section{Conclusion}

We present \tool, a novel approach that leverages LLM-based inference and program trace analysis to compute dynamic data dependencies. \tool achieves superior precision and recall compared to state-of-the-art slicers (e.g., Slicer4J, ND-Slicer), effectively addressing alias inference and removing the need for re-execution. This makes \tool particularly practical for real-world debugging. For future work, we plan to construct a more comprehensive dynamic dependency dataset and expand \tool to support additional programming languages.

\bibliographystyle{ACM-Reference-Format}
\bibliography{main}

\appendix

\clearpage

\def\speciallstcolor{\begingroup\color{blue}}
\def\endspeciallstcolor{\endgroup}

\definecolor{bgprompt}{rgb}{0.95,0.95,1.00}
\definecolor{bgres}{rgb}{1.0,0.95,0.95}

\lstdefinestyle{mypromptstyle}{
    backgroundcolor=\color{bgprompt},   
    commentstyle=,
    keywordstyle=,
    stringstyle=,
    basicstyle=\ttfamily\footnotesize,
    breakatwhitespace=false,
    breaklines=true,
    captionpos=b,
    keepspaces=true,
    numbers=none,
    showspaces=false,
    showstringspaces=false,
    showtabs=false,
    tabsize=2,
    breakindent=0em,
    escapeinside={(*@}{@*)}
}

\lstset{style=mypromptstyle}

\section{Detailed Prompts}

In this section, we present detailed LLM prompts for variable recovery, alias inference, and definition inference. For each task, we provide an example of the prompt and its corresponding response. Within each prompt, the blue text indicates the placeholders that will be replaced with actual values during execution.

\subsection{Variable Recovery}
\label{sec:prompt-varrecovery}

\noindent\textbf{Prompt:}

\begin{lstlisting}
# Background
When executing Java code involving libraries, understanding the internal structure of objects is critical for debugging. Given the "toString" value of a value, your task is to **analyze a Java object** and **generate a expanded JSON representation of internal values** that captures its internal fields relevant to debugging.

During generated JSON representation, you will be given a focal vairable path, e.g. `list.elementData[0].name`, and for objects fields, you should focus on the focal variables and their values, and you can omit the rest fields. But DO NOTICE that, the output should be a JSON object, strictly following the JSON grammar, to omit a field in a JSON object, DO NOT use "...", but just delete the field to keep the JSON valid. For list or array, DO NOT omit any element in LIST. If the focal variable is "#all_fields#", you should ignore the focal variable and return a complete JSON object with the all fields.

The expanded JSON representation should:

- Strictly follow the format `"var_name|var_type": var_value`.
- Reflect inferred values and types for all fields used or affected in the operation.
- Be rooted at the top-level variable (e.g., `set`).
- Wrap your output within a json code block (i.e., ```json ... ```). Do not include any comments inside the json code block.
- If needed, use <thought>xxxx</thought> tags to express your thought process first. The thoughts should be as short as possible.
- If the focal variable contains a deep recursive structure (e.g., a tree or a linked list), you should only expand the first level of the structure. For example, if the focal variable is a tree, you should only expand the root node and its immediate children, but not the entire tree.

# Example

**Focal Variable toString Value:**
(*@\aftergroup\speciallstcolor@*)`{pair={}, }`(*@\aftergroup\endspeciallstcolor@*)

**Focal Variable Type Name:**
(*@\aftergroup\speciallstcolor@*)`java.util.concurrent.atomic.AtomicStampedReference`(*@\aftergroup\endspeciallstcolor@*)

**Related Class Structures:**
(*@\aftergroup\speciallstcolor@*)- `java.util.concurrent.atomic.AtomicStampedReference:{java.util.concurrent.atomic.AtomicStampedReference$Pair pair;}`(*@\aftergroup\endspeciallstcolor@*)


**Focal Variable Path:**
(*@\aftergroup\speciallstcolor@*)`atomicRef.pair.reference`(*@\aftergroup\endspeciallstcolor@*)

**Expected Output:**
(*@\aftergroup\speciallstcolor@*)```json
{
  "atomicRef": {
    "pair|java.util.concurrent.atomic.AtomicStampedReference$Pair": {
      "stamp|int": "1",
      "reference|java.lang.Integer": "42"
    },
    "UNSAFE|sun.misc.Unsafe": "{}",
    "pairOffset|long": "12"
  }
}
```(*@\aftergroup\endspeciallstcolor@*)

## Task

**Focal Variable Name:**
(*@\aftergroup\speciallstcolor@*)`atomicRef`(*@\aftergroup\endspeciallstcolor@*)

**Focal Variable Type Name:**
(*@\aftergroup\speciallstcolor@*)`java.util.concurrent.atomic.AtomicStampedReference`(*@\aftergroup\endspeciallstcolor@*)

**Focal Variable toString Value:**
(*@\aftergroup\speciallstcolor@*)`{pair={}, }`(*@\aftergroup\endspeciallstcolor@*)

**Focal Variable Path:**
(*@\aftergroup\speciallstcolor@*)`atomicRef.pair.reference`(*@\aftergroup\endspeciallstcolor@*)

**Related Class Structures:**
(*@\aftergroup\speciallstcolor@*)- `java.util.concurrent.atomic.AtomicStampedReference:{java.util.concurrent.atomic.AtomicStampedReference$Pair pair;}`(*@\aftergroup\endspeciallstcolor@*)


**Line of Code Containing the Variable:**
(*@\aftergroup\speciallstcolor@*)```java
Integer value = atomicRef.getReference();
```(*@\aftergroup\endspeciallstcolor@*)

**Output:**
\end{lstlisting}

\noindent\textbf{Response:}

\begin{lstlisting}[backgroundcolor=\color{bgres}]
**Output:**

<thought>Focus on the `atomicRef.pair.reference` field and expand its relevant structure while omitting unrelated fields.</thought>

```json
{
  "atomicRef": {
    "pair|java.util.concurrent.atomic.AtomicStampedReference$Pair": {
      "reference|java.lang.Integer": "42"
    }
  }
}
\end{lstlisting}

\subsection{Alias Inference by LLM}
\label{sec:prompt-alias}

\noindent\textbf{Prompt:}

\begin{lstlisting}
<Background>
You are a Java expert, you need to analyze the alias relationships through static analysis. Given a variable and a method call, your task is to identify any alias relationship between (*Set 1*) the listed fields of the given variable and (*Set 2*) the variables involved in the method call and the return value of the method call.

<Example>
Given code:
```list.add(item);```

Given the source code of function calls in the code:
public boolean add(E e) {
modCount++;
add(e, elementData, size);
return true;
}

Variables involved in the line of code:
`list` is of type `java.util.ArrayList`,
`item` is of type `Integer`,

We know that another variable not in the code, `list`, with the following structure:
{"list:java.util.ArrayList":{"elementData:java.lang.Object[]":"[]","size:int":"0"}}

We are interested in the fields `list.elementData.elementData[0]`

Your response should be:
{
"list.elementData.elementData[0]":"item"
}

<Question>
Given code:
(*@\aftergroup\speciallstcolor@*)```AtomicStampedReference<Integer> atomicRef = new AtomicStampedReference<>(initialRef, initialStamp);```(*@\aftergroup\endspeciallstcolor@*)

Given the source code of function calls in the code:
(*@\aftergroup\speciallstcolor@*)public AtomicStampedReference(V initialRef, int initialStamp) {
    this.pair = Pair.of(initialRef, initialStamp);
}(*@\aftergroup\endspeciallstcolor@*)

Variables involved in the line of code:
(*@\aftergroup\speciallstcolor@*)`initialRef` is of type `java.lang.Integer`,
`initialStamp` is of type `int`,
`AtomicStampedReference_instance` is of type `java.util.concurrent.atomic.AtomicStampedReference`,
`atomicRef` is of type `java.util.concurrent.atomic.AtomicStampedReference`,(*@\aftergroup\endspeciallstcolor@*)

(*@\aftergroup\speciallstcolor@*)`initialRef`(*@\aftergroup\endspeciallstcolor@*) has the following fields:
(*@\aftergroup\speciallstcolor@*)`AtomicStampedReference_instance` has the following fields: {"pair":"java.util.concurrent.atomic.AtomicStampedReference$Pair",},(*@\aftergroup\endspeciallstcolor@*)

If a variable has name of format `<TYPE>_instance`, it refers to the instance created by calling the constructor of `<TYPE>`.
If a variable has name of format `return_of_<method_signature>`, it refers to the variable returned by a method call of `<method_signature>`.

We know that another variable not in the code, (*@\aftergroup\speciallstcolor@*)`atomicRef`(*@\aftergroup\endspeciallstcolor@*), with the following structure:
(*@\aftergroup\speciallstcolor@*){"atomicRef|java.util.concurrent.atomic.AtomicStampedReference":{"pair|java.util.concurrent.atomic.AtomicStampedReference$Pair":{"reference|String":"42"}}}(*@\aftergroup\endspeciallstcolor@*)
where
(*@\aftergroup\speciallstcolor@*)this `atomicRef` has the same memory address as `AtomicStampedReference_instance` in the line of code,
this `atomicRef` has the same memory address as `atomicRef` in the line of code,(*@\aftergroup\endspeciallstcolor@*)

We are interested in the fields of this instance: (*@\aftergroup\speciallstcolor@*)`atomicRef`,`atomicRef.pair`,`atomicRef.pair.reference`,(*@\aftergroup\endspeciallstcolor@*)

From the given code, identify all the aliases of this (*@\aftergroup\speciallstcolor@*)`atomicRef`(*@\aftergroup\endspeciallstcolor@*) and the fields in this (*@\aftergroup\speciallstcolor@*)`atomicRef`(*@\aftergroup\endspeciallstcolor@*).

In your response, strictly follow the JSON format. The JSON keys are from the listed fields, JSON values are variables or their fields that are the corresponding aliases of the fields. Do not include explanation.
\end{lstlisting}

\noindent\textbf{Response:}

\begin{lstlisting}[backgroundcolor=\color{bgres}]
{
  "atomicRef": "AtomicStampedReference_instance",
  "atomicRef.pair": "AtomicStampedReference_instance.pair",
  "atomicRef.pair.reference": "initialRef"
}
\end{lstlisting}

\subsection{Definition Inference}
\label{sec:prompt-def}

\noindent\textbf{Prompt:}

\begin{lstlisting}
# definition_inference

## Definition Inference Task

You are a Java expert responsible for analyzing variable assignments. 

### Instructions:
1. You will receive a **target line of code**.
2. You will also be given a **variable name** and the **line of code** where this variable is used.
3. Your objective is to determine whether the target line writes to the specified variable.

### Response Format:
Please provide a clear answer based on your analysis:
- **Answer: <T>** (True) if the target line writes to the variable.
- **Answer: <F>** (False) if it does not.

### Examples:

**Example 1:**
- **h** is an `ArrayList`
- **Target Line:** `h.set(10, "abc");`
- **Variable:** `h.elementData[8]`
- **Usage Line:** `int x = h.get(8);`

**Example Answer 1:**
- **Answer:** <F>  
The result is false because the target line sets the 10th element rather than the 8th element.

---

**Example 2:**
- **d** is a `HashMap`
- **Target Line:** `d.put("ccc", "xyz");`
- **Variable:** `d.table["ccc"]`
- **Usage Line:** `Object r = d.get("ccc");`

**Example Answer 2:**
- **Answer:** <T>  
The result is true because the target line sets the value for the key "ccc" in the hashmap.

---

### Your Turn:
Now, please analyze the following input according to the provided format.
In your response, return <T> for true and <F> for false.

### Question:
**Target Line:**
(*@\aftergroup\speciallstcolor@*)`boolean updated = atomicRef.compareAndSet(expectedRef, newRef, expectedStamp, newStamp);`(*@\aftergroup\endspeciallstcolor@*)

**Function Calls in Source Code:**
(*@\aftergroup\speciallstcolor@*)public boolean compareAndSet(V expectedReference, V newReference, int expectedStamp, int newStamp) {
    Pair<V> current = this.pair;
    return expectedReference == current.reference && expectedStamp == current.stamp && (newReference == current.reference && newStamp == current.stamp || this.casPair(current, Pair.of(newReference, newStamp)));
}(*@\aftergroup\endspeciallstcolor@*)

**Variables Involved:**
(*@\aftergroup\speciallstcolor@*)
Variable: 
`newRef`
Variable Type: 
`java.lang.Integer`
Runtime Value: 
`200`
Variable: 
`newStamp`
Variable Type: 
`int`
Runtime Value: 
`2`
Variable: 
`expectedRef`
Variable Type: 
`java.lang.Integer`
Runtime Value: 
`100`
Variable: 
`expectedStamp`
Variable Type: 
`int`
Runtime Value: 
`1`
Variable: 
`atomicRef`
Variable Type: 
`java.util.concurrent.atomic.AtomicStampedReference`
Runtime Value: 
`{"atomicRef|java.util.concurrent.atomic.AtomicStampedReference":{"pair|java.util.concurrent.atomic.AtomicStampedReference$Pair":{"reference|String":"null"}}}`
(*@\aftergroup\endspeciallstcolor@*)

We know that (*@\aftergroup\speciallstcolor@*)`atomicRef`(*@\aftergroup\endspeciallstcolor@*) has the following structure and value:
(*@\aftergroup\speciallstcolor@*){"atomicRef|java.util.concurrent.atomic.AtomicStampedReference":{"pair|java.util.concurrent.atomic.AtomicStampedReference$Pair":{"reference|String":"42"}}}(*@\aftergroup\endspeciallstcolor@*)
But we don't know which step during the execution modified the value.

**Usage Line:**
(*@\aftergroup\speciallstcolor@*)`Integer value = atomicRef.getReference();`(*@\aftergroup\endspeciallstcolor@*)

(*@\aftergroup\speciallstcolor@*)`atomicRef`(*@\aftergroup\endspeciallstcolor@*) has a field (*@\aftergroup\speciallstcolor@*)`atomicRef.pair.reference`(*@\aftergroup\endspeciallstcolor@*), does the code (*@\aftergroup\speciallstcolor@*)`boolean updated = atomicRef.compareAndSet(expectedRef, newRef, expectedStamp, newStamp);`(*@\aftergroup\endspeciallstcolor@*) directly or indirectly write field (*@\aftergroup\speciallstcolor@*)`atomicRef.pair.reference`(*@\aftergroup\endspeciallstcolor@*)?
In your response, strictly return <T> for true and <F> for false. Briefly explain your answer.
\end{lstlisting}

\noindent\textbf{Response:}

\begin{lstlisting}[backgroundcolor=\color{bgres}]
**Answer:** <T>  
\end{lstlisting}


\end{document}